\documentclass[useAMS,usenatbib]{mn2e}
\usepackage{amsmath,amssymb,wasysym,graphicx,epsfig}

\title[]{\sc{Planetpol} polarimetry of the exoplanet systems 55 Cnc and $\tau$ Boo}

\author[Lucas et al.]{
\thanks{Based on observations made with the William Herschel Telescope operated 
on the island of La Palma by the Isaac Newton Group in the Spanish Observatory
del Roque de los Muchachos of the Instituto de Astrofisica de Canarias.}
P. W. Lucas$^1$ 
\thanks{E-mail: P.W.Lucas@herts.ac.uk.},
J. H. Hough$^1$, J. A. Bailey$^2$, M. Tamura$^3$, E. Hirst$^1$, D. Harrison$^1$\\
$^1$Centre for Astrophysics Research, University of Hertfordshire, College Lane, 
Hatfield AL10 9AB, United Kingdom\\
$^2$School of Physics, University of New South Wales, NSW 2052, Australia\\
$^3$National Astronomical Observatory, Osawa 2-21-2, Mitaka, Tokyo 181, Japan\\}

\begin{document}

\date{Accepted 2008. Received 2008; in original form July 2008}

\pagerange{\pageref{firstpage}--\pageref{lastpage}} \pubyear{2008}

\maketitle

\label{firstpage}

\begin{abstract}
We present very sensitive polarimetry of 55 Cnc and $\tau$ Boo in an attempt to detect
the partially polarised reflected light from the planets orbiting these two stars. 
55 Cnc is orbited by a hot Neptune planet (55 Cnc e) at 0.038 AU, a hot Jupiter
planet (55 Cnc b) at 0.11 AU, and at least 3 more distant planets. The polarisation of 
this system is very stable, showing no sign of the periodic variations that would be 
expected if a short period planet were detected. The measured standard deviation
of the nightly averaged Stokes Q/I and U/I parameters is 2.2$\times 10^{-6}$.
We derive upper limits on the geometric albedo, $A_G$ and planetary radius using Monte Carlo
multiple scattering simulations of a simple model atmosphere. We assume
Rayleigh-like scattering and polarisation behaviour
(scaled by the maximum polarisation, $p_m$ at 90$^{\circ}$) and pressure insensitive extinction. 
Atmospheres in which multiple scattering plays only a small role have an almost linear
relation between polarisation and $A_G$. In this case, the 4$\sigma$ upper limit 
is $A_G<0.13(R/1.2 R_{Jup})^{-2}p_m^{-1}$ for 55 Cnc e. This is most easily explained if 55 Cnc e 
is relatively small, like GJ436b, and therefore not a pure H-He planet. The data do not provide a 
useful upper limit for 55 Cnc b. $\tau$ Boo is orbited by an unusually massive hot Jupiter 
planet. The data show a standard deviation in the night to night average Stokes Q/I and U/I 
polarisation parameters of 5.1$\times 10^{-6}$. The 4$\sigma$ upper limit is
$A_G<0.37(R/1.2 R_{Jup})^{-2}p_m^{-1}$ for $\tau$ Boo b, adopting the fairly well established 
orbital inclination $i$$\sim$40$^{\circ}$. This extends the similar upper limits reported previously
for this planet to longer wavelengths. The fact that the $\tau$ Boo data 
show more scatter, despite the smaller photon noise for this bright star, may be due to the 
spot activity detected photometrically by the {\it MOST} satellite. 
These results contrast markedly with the recent claim of a 3$\sigma$ detection of a periodic 
polarisation signal from HD189733 with amplitude P$=2\times 10^{-4}$, attributed to the planet 
HD189733~b. 
\end{abstract}


\section{Introduction}

	Since the seminal discovery of 51 Peg b (Mayor \& Queloz 1995),
$\sim$300 extrasolar planets have been found (e.g. Butler et al.2006, 
Schneider 2008). Approximately one third of
these orbit very close to their central star, with semi-major axes, $a<0.1$~AU
(Schneider 2008). The number of short period planets is increasing rapidly
with the growing success of transit based searches such as SuperWASP (Pollacco
et al.2006) and HATNet (Bakos et al.2004), which are even more strongly biased
towards such detections than the radial velocity method.

These close in planets offer the best opportunity to study the physical 
characteristics of the planets, as opposed to their orbital parameters.
The combination of transit data and radial velocity data yields a precise 
planetary mass, size and density. Furthermore, recent studies with the
{\it Spitzer} Space Telescope (e.g. Knutson et al.2007; 2008; Burrows et al.2007;
Tinetti et al.2007) and the {\it Hubble} Space Telescope (e.g. Pont et al.2008; Swain et al.2008; 
Lecavelier des Etangs 2008) have obtained a great deal of information about the 
chemical composition and pressure temperature profiles of the atmospheres of the 
two brightest transiting ``hot Jupiter'' planets, HD189733~b and HD209458~b.
These studies used infrared photometry and spectroscopy of radiation emitted
by the planets and optical spectrophotometry of the transit. All required 
exceptionally high signal to noise data for the integrated light of the
star and planet. These results build on the more limited data provided by earlier
space based studies, e.g. Charbonneau et al.(2002); Harrington et al.(2006).

While space-based observations have now produced high quality data on the 
atmospheres of these two planets, ground based observations have yet to produce
a confirmed direct detection of an extrasolar planet. Neither space based nor 
ground based efforts have yet detected an extrasolar planet in reflected light.
The MOST satellite has provided strong upper limits on the reflected light
from HD209458~b (Rowe et al.2007) and ground based searches have also produced meaningful
upper limits for a few planets (e.g. Charbonneau et al.1999; Leigh et al.2003; Winn et al.2008; 
Snellen 2005; Snellen \& Covino 2007). Recently, Berdyugina et al.(2008) have reported 
a possible detection of reflected light from HD189733~b at low signal to noise,
using ground based polarimetry with a conventional instrument employing 
slow modulation. However, the reported fractional polarisation is so large 
($P \sim 2 \times 10^{-4}$) that it exceeds that expected for a perfectly
reflective planet (i.e. a Lambert sphere with Rayleigh scattering polarisation properties) 
with the size indicated by the transit data, see Seager, Whitney \& Sasselov (2000).

\begin{table*}
\begin{minipage}{100mm} 
    \caption{Orbital parameters}
    \begin{tabular}{lccccc}
  \hline
Planet & $m.sin(i)$ (M$_{Jup}$) & a (AU) & Period (d) & e & i \\ \hline
55 Cnc e &   0.034    	      &  0.038   & 2.817  & 0.07$\pm$0.06 & $\sim55^{\circ}$     \\
55 Cnc b &   0.824    	      &  0.115   & 14.652  & 0.014$\pm$0.008 & $\sim55^{\circ}$  \\
$\tau$ Boo b &  4.13   	      &  0.0481   & 3.312  & 0.023$\pm$0.015 & $\sim40^{\circ}$  \\
 \hline
\end{tabular}
\end{minipage}
\end{table*}

Polarimetry is nonetheless a promising method to detect extrasolar planets 
in reflected light. The basic principle is that reflected light is partially 
polarised (Seager et al.2000; Stam, Hovenier \& Waters 2004; Stam et al.2006; Stam 2008)
whereas the direct light emitted by stellar photospheres has negligible 
linear polarisation. Kemp et al.(1987) measured an upper limit on the 
integrated linear polarisation from the solar disc of $P<2\times 10^{-7}$ 
in the V band. Polarimetry therefore circumvents the contrast problem 
associated with extrasolar planets and their central stars. It can be used
to detect planets that are spatially unresolved from the central star by 
searching for changes in the polarisation that have the same period as the planet's 
orbit (Seager et al.2000). In this case the time varying polarisation due to the
planet will usually be superimposed on a constant polarisation that is caused by dichroic 
extinction in the interstellar medium along the line of sight to Earth. The interstellar 
polarisation of stars in the local ionised bubble within $\sim$100~pc of the sun is very small 
(typically P$\sim$10$^{-5}$-10$^{-6}$, see Hough et al.2006) and is assumed to be constant
during the period of an observing run, see $\S$4.4.
Alternatively, polarimetry can be used to assist in detecting planets that are spatially 
resolved from the central star by very high resolution imaging. This approach is
be attempted with new instruments such as Hi-CIAO on Subaru, {\sc NICI} and {\sc GPI} on 
Gemini and {\sc SPHERE} on the VLT.

In this paper we have adopted the first approach, using an aperture polarimeter
to observe exoplanet systems in which the planet cannot be spatially resolved in the
forseeable future. Hough et al.(2006) described the {\sc Planetpol} instrument, which was 
specifically designed to achieve a sensitivity to fractional linear
polarisations of order 10$^{-6}$ by using a fast modulator system, similar
in principle to that described by Kemp et al.(1987) and references therein. 
Polarimetry with fast modulators 
(which induce a rapid periodic variation of the relative retardance of electromagnetic
vibrations in orthogonal planes) permits this very high sensitivity by separating 
the polarised component of the incident radiation field from the unpolarised 
component, converting the former into a high frequency signal in the 
time domain. 
This means that very small fractional polarisations can be measured without
the need for very precise flux measurements with the detector. 
{\sc Planetpol} achieves photon noise limited performance for bright exoplanet systems
so the main limitation on the polarisation sensitivity is the need to gather in excess
of 10$^{12}$ photons in order to measure fractional polarisations as low
as 10$^{-6}$.  The fast modulation is performed by Photoelastic Modulators
(PEMs) operated at 20 kHz, which is fast enough to remove any effects of 
the Earth's atmosphere.

By contrast, conventional polarimeters for night time astronomy employ 
slow modulation, rotating a waveplate with a fixed retardance of 
0.5 wavelengths to several different position angles. Measurement of
fractional polarisations of order $10^{-6}$ with such a system would
require a precision of order $10^{-6}$ in the measured flux on the 
detector, which would be very challenging. In practice such polarimeters
rarely achieve a sensitivity better than $P=10^{-4}$.

While the recent space based measurements have revealed much about the atmospheres of two 
transiting planets the energy balance of these systems cannot yet be understood without 
measurement of their albedos. Most of the space based measurements have been obtained during 
transits, which only sample the upper layers of the atmosphere, not the principal
reflecting layer. Even if space based spectroscopy is obtained outside transit events, 
it is difficult to infer the presence or absence of dust particles and their contribution to the 
radiative transfer without polarisation data. Successful polarimetric detection of exoplanets 
could address these questions and permit determination of the inclination of the planetary orbit 
(transiting systems are not required). For example, Hansen \& Hovenier (1974) used ground-based
aperture polarimetry to determine that sulphuric acid clouds, with $\sim$1~$\mu$m-sized droplets,
dominate the reflection of sunlight from Venus.

Here we describe the results of {\sc Planetpol} polarimetry of two exoplanet
systems, 55 Cnc and $\tau$ Boo. The orbital parameters of their potentially
detectable inner planets are listed in Table 1, taken from Butler et al.(2006) and 
Fischer et al.(2008) (with the exception of the inclinations, see below). These systems 
were selected on the basis that their bright I band magnitudes (brighter than any of 
the known transiting planets) and the short orbital period of their closest 
planetary companions made them good prospects to produce a detectable polarised flux.

The 55 Cnc system (HR3522) is a wide stellar binary, consisting of a G8IV-V primary 
(Baliunas et al.1997) with Cousins I mag=5.1 and an M4 secondary (Patience et al.2002; 
Eggenberger et al.2004; 
Raghavan et al.2006). The primary is now believed to be orbited by 5 planets (Fischer et al.2008; 
Butler et al.1997, Marcy et al.2002; McArthur et al.2004) and the inclination of the planetary 
orbits has some constraints
from high precision astrometry of reflex motion caused by the outermost planet (McArthur et al.2004).
It has an R$\prime$$_{HK}$ index of -4.97 (Soderblom 1985; Baliunas et al.1996) 
and a rotation period of 44 days (Soderblom 1985). The innermost planet, 55 Cnc e, 
is a ``hot Neptune'' and the next planet out, 55 Cnc b, is a hot Jupiter. Both planets could 
reasonably be expected to produce a detectable signal, although the relatively large orbit
of 55 Cnc b (see Table 1) would require it to have a high albedo and a large radius, and the 
Neptune mass planet 55 Cnc e might also require a high albedo, depending on the size of the planet.

The $\tau$ Boo system (HR5185) is also a wide stellar binary composed of an F7IV-V primary 
(Baliunas et al.1997) with Cousins I mag=4.0 and an M2V secondary (Raghavan et al.2006). The 
primary is orbited by 
one of the most massive known hot Jupiter planets (Butler et al.1997), which is commonly
referred to as $\tau$ Boo b. Baliunas et al.(1997; 1996) found the star to be photometrically 
stable at the millimagnitude level, with an R$\prime$$_{HK}$ index of -4.73. However, recent
space based photometry by MOST has found that the stellar flux is variable at the millimagnitude
level (Walker et al.2008). The period is consistent with the 3.3 day planetary orbit, suggesting 
that there is magnetic activity on the stellar surface that is linked to the planet. The changes
in flux appear to be due mainly to a single star spot. However, the amplitude of the changes
varies on long timescales, in some years apparently corresponding to a bright spot rather
than a dark spot. $\tau$ Boo is believed
to be a unique case where the massive planet has tidally spun up at least the outer layers
of the star to have the same rotation period as the planet's orbit. The inclination of the
stellar rotation axis can be determined from the measured photometric period and the
$v sin(i)$ Doppler broadening of stellar lines as $i$ $\approx$ 40$^{\circ}$, e.g. Leigh et al.(2003). 
Recent mapping of the magnetic field using (circular) spectropolarimetry (Catala et al.2007) also 
finds a similar intermediate inclination.
Assuming that the planetary orbit shares this inclination, this allows the mass of the planet to 
be determined as $\sim$ 6-7~M$_{Jup}$.

\section{Observations}

All observations were made with {\sc Planetpol} mounted at the Cassegrain focus of
the 4.2-m William Herschel Telescope at the Roque de los Muchachos Observatory on La Palma,
one of the Canary Islands.

\subsection{55 Cnc}

\begin{figure}
\begin{center}
\begin{picture}(200,285)

\put(0,0){\includegraphics{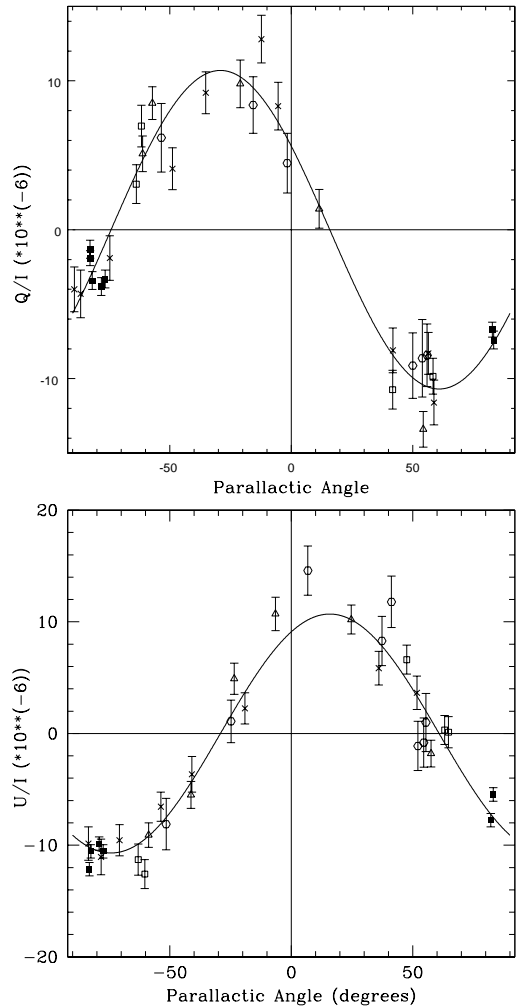}}

\end{picture}
\end{center}
\vspace{3.2cm}
\caption{The fit to the telescope polarisation (TP) in the Februry 2006 observing run, plotted
in instrumental coordinates.
The data for the nearby bright stars that are used to construct the fit are plotted
with different sysmbols for each star, after subtraction of the constant intrinsic Q/I
and U/I values given in Table 2 (see Hough et al.2006 for a full explanation of the 
fitting procedure). The quality of the fit illustrates the polarisation stability of normal stars.
Ten outlying measurements have been excluded from both the plot and the fit (see $\S$2.3).
{\it (upper panel)} Q/I; {\it (lower panel)} U/I. The error
bars are the internal random errors on the measurements.}
\end{figure}

55 Cnc was observed on 15-20 February 2006. The OG590 longpass filter was used in each case, leading 
to a central wavelength near 780~nm and a broad response between 590 and 920~nm. The long wavelength
limit is determined by the response of the single element Avalanche Photodiode Detectors in 
{\sc Planetpol}. The PEMs were 
set to an amplitude of 0.5 $\lambda$ at 780~nm. The 55 Cnc system and the nearby low 
polarisation stars used to calibrate the telescope polarisation (TP) were all observed with a
5 arcsec diameter aperture.
The observing procedure, system response and data reduction is described in Hough et al.(2006). 
The instrument simultaneously measures the total flux and the polarised flux within the aperture,
providing Stokes I and Q data or Stokes I and U data, depending on the orientation of the instrument.
The reduction script provides the corresponding Q/I or U/I value. The Stokes Q and U parameters are 
measured in the instrumental system and are later converted to the equatorial system. 
Integration times of 24 minutes on source were used for the individual Stokes Q and U
data points for 55 Cnc. This time was split into 8 steps of 180~s each by the second 
stage chopping procedure, in which the orientation of the analyser and detector assemblies 
alternates between $+45^{\circ}$ and $-45^{\circ}$ relative to the PEMs, thereby reversing the 
sign of the measured polarisation in order to aid removal of systematic effects.  With 
overheads for second stage chopping and readout the elapsed time for a Stokes Q or U data 
point was $\approx 28$ minutes. 

The nearby stars used to calibrate the TP were HR2421, HR2990, HR4534, HR4540 and HR4295. 
They were observed in the same manner as $\tau$ Boo, except that the time on source
for individual Stokes Q and U measurements was 12 minutes, comprising only 4 steps.   
These stars all have low polarisation and are brighter than 55 Cnc, so this shorter integration
time is sufficient to calibrate the TP with the requisite precision (Hough et al.2006).
Their properties and mean fitted 
polarisations are summarised in Table 2. The individual measurements, after subtracting the fitted TP
and the instrumental polarisation (IP), are listed in Table 3 to illustrate their stability.
Note that the errors quoted in Table 3 are the internal random errors associated with each 
measurement. These typically underestimate the true errors slightly, see $\S$2.3.

\begin{table*}
\begin{minipage}{100mm} 
    \caption{Stars used to calibrate Telescope Polarisation during Feb 2006. Average polarisation 
measurements are given in the equatorial system.}
    \begin{tabular}{lccccccc}
  \hline
Star   & Spectral Type & Distance & I$_c$ & V-I$_c$ & Number of & $<$Q/I$>$ & $<$U/I$>$ \\
       &               &  (pc)    &       &       & obs (Q/U) & ($\times$10$^{-6}$) & ($\times$10$^{-6}$) \\ \hline
HR2421 &  A0IV         &  31.2    & 1.89  & 0.04  &  (5/6) & -1.49$\pm$1.21 & -7.01$\pm$1.14 \\
HR2990 &  K0III        &  10.3    & 0.19  & 0.97  &  (7/6) & -12.82$\pm$0.84 & -12.04$\pm$0.92 \\
HR4534 &  A3V          &  11.1    & 2.04  & 0.10  &  (4/5) &  3.85$\pm$1.22 & -6.01$\pm$1.28 \\
HR4295 &  A1V          &  24.3    & 2.32  & 0.02  &  (10/8) &  5.00$\pm$0.84 & -8.19$\pm$0.99 \\
HR4540 &  F8V          &  10.9    & 2.98  & 0.61  &  (6/8) &  3.26$\pm$1.39 & -0.12$\pm$1.14 \\
\end{tabular}
\end{minipage}
\end{table*}

\begin{table*}
\begin{minipage}{100mm} 
    \caption{Individual Q/I and U/I measurements (in units of 10$^{-6}$) given in the instrumental 
system for the TP calibration stars used in Feb 2006.}
    \begin{tabular}{lcccccccc}
  \hline
Star & Date & Time & Q/I & $\sigma$  & Date   & Time & U/I & $\sigma$ \\ \hline
HR2421 & Feb 15  & 20:07   & -6.5  & 1.3 & Feb 15  &  20:25   & 2.0   & 1.2  \\
HR2421 & Feb 17  & 20:18   & -4.2  & 1.2 & Feb 17  & 21:08    &  0.7  & 1.3  \\
HR2421 & Feb 18  & 21:32   & -7.4  & 1.7 & Feb 18  & 21:46    & 5.3   & 1.6  \\
HR2421 & Feb 18  & 23:19   & -10.2 & 1.3 & Feb 18  & 23:34    & -1.5  & 1.3  \\
HR2421 & Feb 19  & 22:01   & -7.2  & 1.4 & Feb 19  &  21:24   & 4.9   & 1.5  \\
   -   &  -      &  -  &   -   &  -      & Feb 19  & 22:16    & 1.9   & 1.3   \\ \hline  
HR2990 & Feb 15  & 21:56   & -10.0 & 0.5 & Feb 15  &  22:33  & 13.5  & 0.6  \\
HR2990 & Feb 16  & 00:14   & -11.2 & 0.6 & Feb 16  &  00:29  & 12.3  & 0.6  \\
HR2990 & Feb 16  & 19:54   & -13.9 & 0.6 & Feb 16  & 20:08  & 16.8  & 0.6  \\
HR2990 & Feb 16  & 21:53   & -9.3  & 0.6 & Feb 16  & 22:28  & 10.2   & 0.6  \\ 
HR2990 & Feb 17  & 21:28   & -11.9 & 0.6 & Feb 17  & 21:42    & 16.5  & 0.6  \\ 
HR2990 & Feb 18  & 00:36   & -10.2 & 0.5 & Feb 18  & 00:50    & 13.5  & 0.6  \\
HR2990 & Feb 20  & 20:00   & -13.8 & 0.6 & Feb 20  & 20:15    & 11.3  & 0.6  \\
HR2990 & Feb 21  & 03:25    & -5.4  & 0.6 & Feb 21  &  03:08   & 15.7  & 0.7  \\ \hline
HR4534 & Feb 16  & 00:49    & -3.5  & 1.5 & Feb 16  & 01:04    & -5.7  & 1.4  \\ 
HR4534 & Feb 16  & 05:24    & -5.3  & 1.3 & Feb 16  & 05:40    & -1.3  & 1.4  \\ 
HR4534 & Feb 17  & 00:12    & -7.0  & 1.4 & Feb 17  & 00:27    & -3.4  & 1.5  \\ 
HR4534 & Feb 17  & 05:57    &  2.0  & 1.3 & Feb 17  & 06:12    &  1.3  & 1.4  \\ 
HR4534 & Feb 18  & 04:08    & -8.8  & 1.4 & Feb 18  & 04:22    & -3.9  & 1.4  \\ 
HR4534 & Feb 19  & 06:46    &  2.0  & 1.9 & Feb 20  & 06:44    & -4.1  & 1.5  \\ \hline
HR4295 & Feb 16  & 03:47    & -9.8  & 1.5 & Feb 16  & 04:02    & -3.9  & 1.7  \\
HR4295 & Feb 16  & 06:59    & -6.9  & 1.6 & Feb 17  & 01:04    & -7.1  & 1.6  \\
HR4295 & Feb 17  & 00:49    & -8.1  & 1.6 & Feb 17  & 04:35    & -2.9  & 1.4  \\
HR4295 & Feb 17  & 04:21    & -13.0 & 1.5 & Feb 17  & 06:28    & -4.2  & 1.6  \\
HR4295 & Feb 17  & 06:43    & -8.2  & 1.7 & Feb 18  & 00:16    & -4.2  & 1.6  \\
HR4295 & Feb 18  & 00:02    & -6.0  & 1.5 & Feb 18  & 05:30    & -3.4  & 1.5  \\
HR4295 & Feb 19  & 23:46    & -9.5  & 1.6 & Feb 20  & 00:02    &  2.2  & 1.6  \\
HR4295 & Feb 20  & 02:24    & -7.3  & 1.7 & Feb 20  & 02:39    & -11.7 & 1.6  \\
HR4295 & Feb 20  & 05:38    & -10.3 & 1.6 & Feb 20  & 05:53    & -5.1  & 1.7  \\
HR4295 & Feb 21  & 02:35    & -4.2  & 1.7 & Feb 21  & 02:50    & -6.1  & 1.5  \\ \hline
HR4540 & -  &  -   & -  &  -  & Feb 18  & 04:43  & -2.7  & 2.4  \\
HR4540 & -  &  -   & -  &  -  & Feb 18  & 04:58  &  2.2  & 2.5  \\
HR4540 & Feb 19  & 00:05    & -1.3  & 2.5 & Feb 19  & 00:20  & -3.9  & 2.5  \\
HR4540 & Feb 19  & 05:41    &  0.6  & 2.4 & Feb 19  & 05:55    & -7.9  & 2.4  \\
HR4540 & Feb 19  & 06:11    &  1.6  & 2.8 & Feb 19  & 06:26    & -4.4  & 2.8  \\
HR4540 & Feb 20  & 02:58    & -2.1  & 2.2 & Feb 20  & 03:16    &  1.6  & 2.4  \\
HR4540 & Feb 20  & 06:27    &  2.0  & 2.3 & Feb 20  & 06:12   & -6.7  & 2.4  \\
HR4540 & Feb 21  & 02:17    & -1.6  & 2.1 & Feb 21  & 02:02    & -3.8  & 2.1  \\ \hline
\end{tabular}
\end{minipage}
\end{table*}

{\sc Planetpol} has a separate optical channel which is used to measure the polarised flux
from the night sky at a distance of $\sim$14 arcminutes from the target, in order to subtract 
this from the flux measured in the science channel.
The February 2006 run occurred during bright time and for a significant fraction of the time 
there was a polarised flux from the sky (possibly enhanced by reflection of moonlight 
from very thin cirrus) that was of the same order as the polarised flux from 55 Cnc in the raw 
data. The stability of the subtracted measurements (see $\S$3.1) indicates that 
this procedure worked well. There is much less photon noise in the sky measurements than in 
measurements of bright stars since it is a faint but highly polarised radiation field. 
Consequently the sky subtraction does not make a significant contribution to the total 
error budget.

The position angle and degree of polarisation were calibrated using the polarisation
standards HD43384, HD154445, and HD21291 (Serkowski 1974)

\subsection{$\tau$ Boo}

$\tau$ Boo was observed in four observing runs on 24-26 April 2004, 25-30 April 2005, 7-8 May 2005
and 15-20 Feb 2006. The 2006 observing run was primarily devoted to 55 Cnc but a pair of Stokes
Q/I and U/I observations of $\tau$ Boo were taken at the end of each night after 55 Cnc had set.
The observing set up was as described above for 55 Cnc. The nearby bright stars used to calibrate the 
TP in the 2005 runs were HR5854, HR4534, HR4932 and HR 5435. Their physical characteristics and 
mean polarisations are given in Table 4 of Hough et al.(2006) and the individual measurements are listed 
in Table 2 of Bailey et al.(2008). In the 2004 run the stars used were HR4540, HR 5854
and HR5793. We do not list the individual measurements for these three stars, which were too few to 
carefully investigate their stability. However, the measurements are shown graphically in Figure 6 of 
Hough et al.(2006).

Data taken on 3-7 May 2005 were contaminated by a Saharan dust event above the observatory 
(Bailey et al.2008; Ulanowski et al.2007). This produced a polarisation of order 
10$^{-5}$ at large zenith distances which gradually declined throughout that period. We
exclude the data taken on 3-6 May from our analysis but we include the $\tau$ Boo
observations from 7 May, since this was at the end of the dust event 
and the observations were taken at zenith distances $\le$20$^{\circ}$, which reduced the effect
on fractional polarisation to a level below the measurement error. The $\tau$ Boo data 
for that night are therefore useable but we have conservatively increased the uncertainty on the 
measurements by a factor of $\sqrt{2}$. Data 
from 7 May were not included in the calibration of TP. The data from the nights of 7 and 8 May provide 
useful repeat coverage of the same orbital phase as 27 and 28 April respectively (after an 
interval of 3 orbits).

The position angle and degree of polarisation was calibrated using the polarisation
standards HD198478 and HD187929 in 2004. In April 2005 the standards HD187929, 
HD147084, HD198478 and HD154445 were used. In May 2005 the standards HD183143 and 
HD198478 were used. All were taken from Serkowski (1974).

\subsection{Stability of Planetpol measurements}

The random errors on measurements with {\sc Planetpol} can be determined from the scatter
of the residuals to TP fits, such as that shown in Figure 1. In Hough et al.(2006) we reported 
that these errors tended to be larger than the internal errors (determined from the standard 
deviation of the polarised flux within each measurement) by a factor of 1.6 to 1.8. In that 
work we reported that this factor did not show an obvious dependence on the flux from the star.
However, the data on the five bright nearby stars in Table 3 include more repeat measurements of 
each star than previously, including the very bright star HR2990 ($\beta$ Gem). This star has I mag=0.2,
leading to a typical internal error on Q/I or U/I measurements of only 6$\times$10$^{-7}$.
The residuals to the TP fit for this star were more than three times this value on several occasions.
Furthermore, some measurements of the other four bright stars deviated by $>$3$\sigma$ from the fit.
Such measurements comprise 16\% of the total (12/75) for these five stars. Ten of these outlying 
measurements deviated from the TP fit by $>$3$\times$10$^{-6}$ and were excluded from the fitting process. 
These discrepancies appear to be less common among the fainter stars, for which larger fractional errors 
in Q/I and U/I are expected, in proportion to photon noise. No 3$\sigma$ errors were found for the 
faintest of the five stars, HR4540 (I mag=3.0), and the measurements of 55 Cnc (I mag=5.1) are all 
very stable and consistent with the expected scatter if no planet were detected, see $\S$3.1.

It therefore seems likely that there is another source of error in addition to the random 
error due to photon noise (see $\S$2.4). Inspection of the discrepant measurements
showed that they are usually associated with disagreements in the measurements in the two beams in the 
science channel. These occasional differences in the two beams emerging from the Wollaston prism
were previously reported in Hough et al.(2006) but their origin is not understood. We found
no correlation between the discrepancies and the flux in the sky channel. Since deviations
from the TP fit as high as 7-8$\times$10$^{-6}$ have occasionally been seen in bright stars
we surmise that this additional source of error is non-Gaussian and has a broader distribution 
of values. We conclude that more than one measurement of a science target should be taken whenever
possible in order to guard against erroneous data. Even when obviously outlying measurements are 
excluded, the additional error source appears to impose a limit on the sensitivity of {\sc Planetpol} 
at a level slightly below 1$\times$10$^{-6}$, see Table 2.

\begin{figure}
\vspace{-7.2cm}\includegraphics[scale=1.0,angle=0]{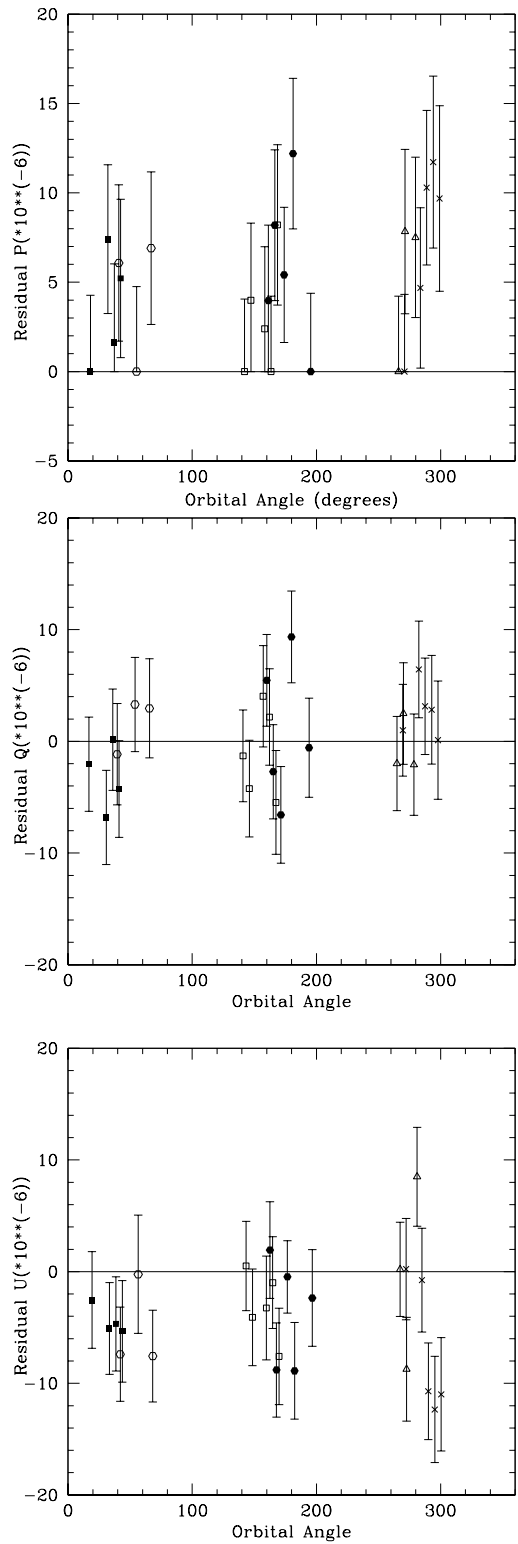}
\vspace{-2.2cm}
\caption{Polarisation of the 55 Cnc system as a function of the orbital angle, $\phi$,
of 55 Cnc e. Minimum illumination occurs at $\phi$=180$^{\circ}$. The polarisation 
of the telescope and the instrument have been subtracted. The data are plotted from $\phi$=0 to 
360$^{\circ}$, as opposed to the usual -180 to 180$^{\circ}$, in order to avoid splitting the data 
near 180$^{\circ}$. 
{\it (top)} fractional polarisation (P=$\sqrt{Q^2+U^2-\sigma^2}/I)$); {\it (middle)} Q/I; 
{\it (bottom)} U/I. The data show no significant sign of variability within nights or between
nights. Data points are assigned a different symbol for each of the six nights 
of observation.}
\end{figure}

\section{Results}
\subsection{55 Cnc}

The polarisation measurements for 55 Cnc are shown in Table 4 and Figure 2 (in the equatorial system), 
as a function of the orbital angle ($\phi$) of 55 Cnc e. The angle $\phi$=180$^{\circ}$ corresponds to
minimum illumination of the observed hemisphere of the planet. The TP and the IP have 
been subtracted from the measurements, leaving only the astronomical signal, which we 
refer to as the residual polarisation. The data from different nights are plotted with 
different symbols. Individual nights cover only a small range in orbital angle so they can be 
usefully averaged. We have used the standard formula P=$\sqrt{Q^2+U^2-\sigma^2}/I)$, where 
any negative P$^2$ values are set to zero. In general the internal uncertainties are very similar 
for the Q and U data so we have simply taken an average of the two for the uncertainty in P. 
The uncertainties in TP and IP make a negligible contribution to the total uncertainty, $\sigma$, of 
approximately 4.3$\times10^{-6}$ on each measurement that is used in the above formula. 

The data cover three separated sections of the orbit of 55 Cnc e, each with a width of 
$\sim 50^{\circ}$. Since the orbital period is close to 3 days the observations on 18 to 
20 February 2006 repeated and broadened the orbital phase coverage of the data taken on 15 to 17 
February. Allowing for the uncertainties, the average Q/I, U/I and P values are constant,
i.e. the values are similar in the three sections of the orbit and the data from
the two orbits agree well. The average values of Q/I, U/I and P are 
$\approx$ 0.2, -4.1 and 4.9 respectively, in units of $\times10^{-6}$. These very small values indicate 
that there is very little interstellar polarisation toward this nearby system (d=13~pc).
This is consistent with the results for the nearby stars given in Table 2 and Hough et al.(2006): 
interstellar polarisation (which is caused by dichroic extinction by aligned dust grains) is of 
order $10^{-6}$ to $10^{-5}$ for stars within $\sim30$~pc.

The polarised flux from the night sky was detectable and comparable to the polarised flux 
from the 55 Cnc system for approximately half of the measurements, when the moon was 
above the horizon and not far from full phase. However, the consistency of the measurements
indicates that the sky polarisation was generally well measured and subtracted. 

\subsection{$\tau$ Boo}

All the polarisation measurements for $\tau$ Boo are shown in Figure 3(a-c) and Table 5 (in the equatorial
system) as a function of the orbital angle of $\tau$ Boo b, with TP and IP subtracted. Again 
$\phi$=180$^{\circ}$ corresponds to minimum illumination of the observed hemisphere.
Data taken on different nights during the 2004 and 2005
observing runs are shown with different symbols. The 2006 data comprise just one datum per night for 
six nights and are all shown as small filled circles. The error bars are based on the internal 
standard error of $\approx 2.5\times10^{-6}$ for each measurement, which is mostly due to photon noise. 
The much smaller uncertainties in TP and IP make very little contribution to the total errors.
The average standard deviation of the $\tau$ Boo data within individual nights 
is only $\approx$16\% higher than the internal standard error. This indicates that the plotted error 
bars are appropriate and there is no measurable variability in the polarisation of the $\tau$ Boo 
system on timescales of a few hours. It also tends to confirm our conclusion in $\S$2.3 that photon
noise usually dominates the error budget at this level of precision.

\begin{figure}
\vspace{-7.6cm}\includegraphics[scale=1.0,angle=0]{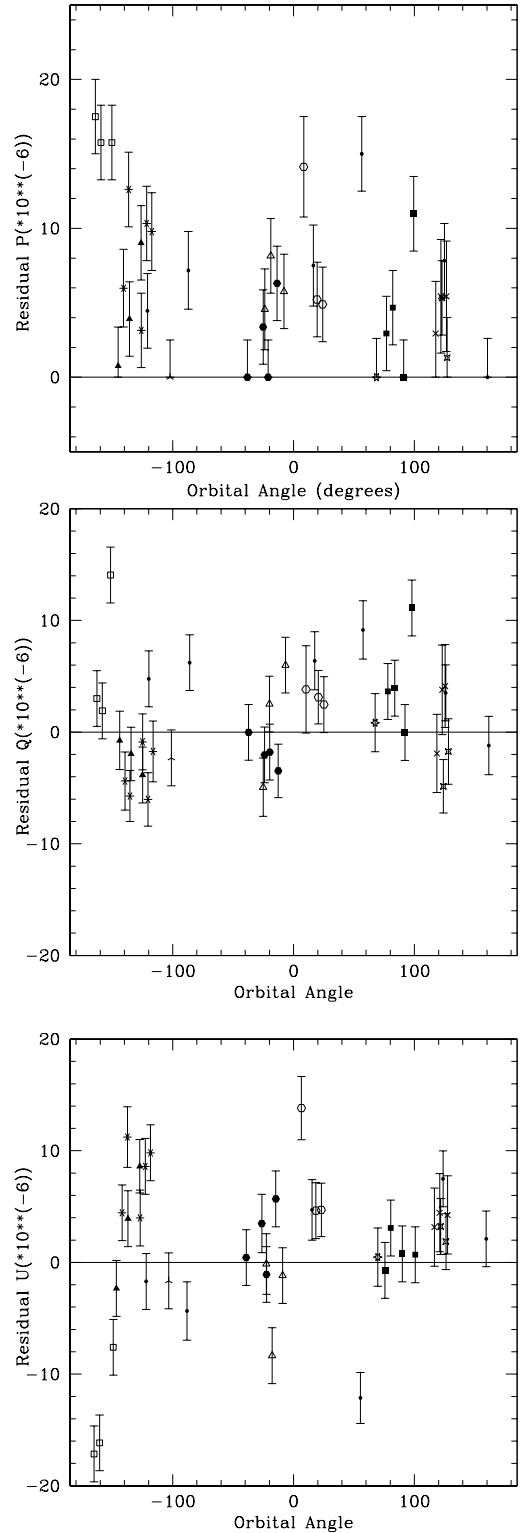}
\vspace{-2.4cm}
\caption{Polarisation of the $\tau$ Boo system as a function of the orbital angle
of $\tau$ Boo b. The polarisation of the telescope and the instrument have been subtracted.
{\it (top)} fractional polarisation, P; {\it (middle)} Q/I; 
{\it (bottom)} U/I. Data taken on each night have similar orbital angles. The 2004 and 2005 data have 
a unique symbol for each night. The 2006 observations had only 1 datum per night and are plotted with small 
filled circles. Overall, the observations show good agreement within each night but
there is a somewhat larger scatter between nights than would be expected from a constant
interstellar polarisation. There is no sign of periodic variability at the 3.3 day period of the
planet.}
\end{figure}

\begin{figure}
\vspace{-7.6cm}\includegraphics[scale=1.0,angle=0]{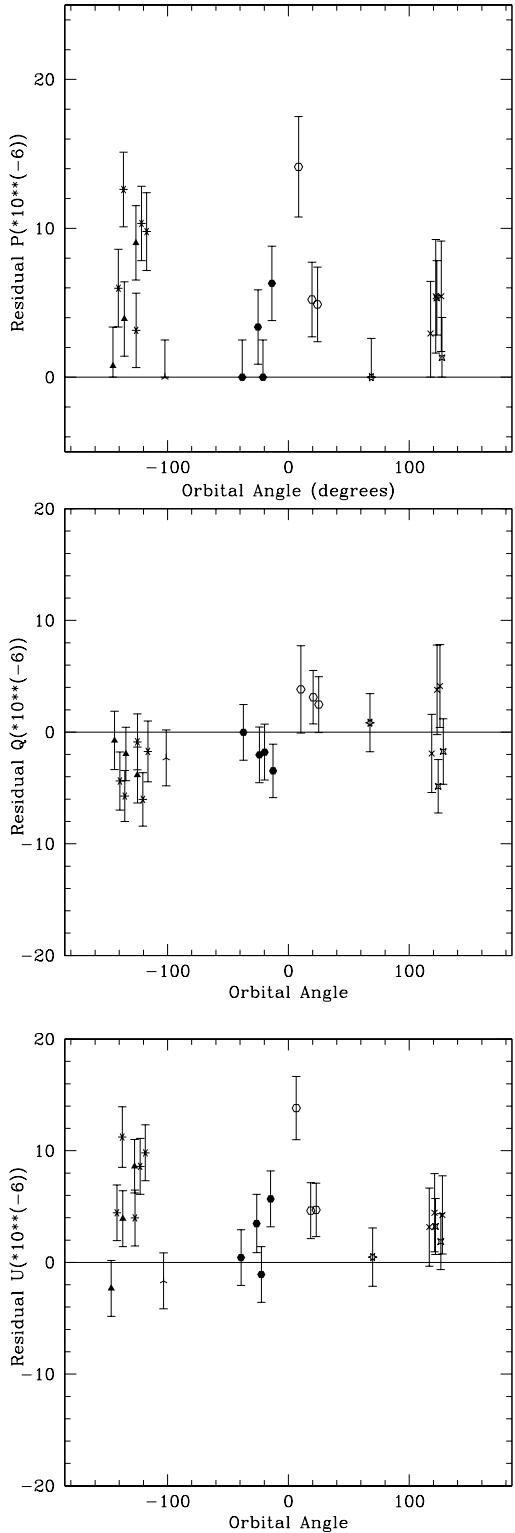}
\vspace{-2.4cm}
\caption{Polarisation of the $\tau$ Boo system in April-May 2005 as a function of the orbital angle
of $\tau$ Boo b. The polarisation of the telescope and the instrument have been subtracted.
{\it (top)} fractional polarisation, P; {\it (middle)} Q/I; 
{\it (bottom)} U/I. Data taken on each night have similar orbital angles. A different symbol is 
used for each night. These plots show less scatter than the plots of all the data in Figure 3.}
\end{figure}

Overall, the data show somewhat more scatter than might be expected from the error bars, and certainly
more than the data for 55 Cnc.
We note that the Q/I and U/I data in Figures 3(b-c) are the measured independent quantities
and that they should measure any changes in polarisation state more clearly than P.
For example, the well sampled region between -165$^{\circ}$ and -115$^{\circ}$ includes
data taken on 26 April 2004, 28 April 2005 and 8 May 2005. In units of $10^{-6}$ the average U/I 
values for these nights are $-13.6\pm3.0$, $3.4\pm3.2$ and $7.6\pm1.4$ respectively (where the 
uncertainties are derived from the scatter in the measurements for each night). The first and last
of these measurements differ from their mean value (-0.9$\times$10$^{-6}$) 
by 4$\sigma$ and 6$\sigma$ respectively. The average Q/I measurements for these nights, in the same 
units, are 6.3$\pm$3.9, -2.2$\pm$1.4, -3.7 $\pm$1.1 respectively. (For the latter two values 
the uncertainties are calculated from the internal photon-noise dominated uncertainties, since the 
scatter in the measurements happens to be smaller). The Q/I measurements lie closer to their mean, 
0.1$\times$10$^{-6}$, but the last measurement is 3$\sigma$ below it.

The data taken in 2005 show less scatter than the full dataset. We plot the 2005 subset of the
data separately in Figure 4(a-c).
The standard deviations of the Q/I and U/I nightly averages for those nights in 2005 with more
than one datum per night are 2.8$\times$10$^{-6}$ and 2.5$\times$10$^{-6}$ respectively.
When we include all nights from 2004 and 2005 with more than one datum the standard deviations
increase to 3.6$\times$10$^{-6}$ in Q/I and 6.5$\times$10$^{-6}$ in U/I. The large standard
deviation in U/I is due in part to the two very negative readings at orbital angle,
$\phi$$\approx$-160$^{\circ}$, taken on 26 April 2004. However, the 2006 data also show a large
standard deviation of 7.0$\times$10$^{-6}$ in U/I, so the large scatter seen in Figure 3(a-c) cannot
be readily attributed to just one or two outlying readings.

Despite these signs of variability in the polarisation of $\tau$~Boo there is no sign of the peaks 
and troughs in Q/I or U/I predicted by the models shown in Figure 6(a-c) that would 
be most likely if reflected light from an extrasolar planet had been detected (see $\S$4.2). The average 
values of $|Q/I|$ and $|U/I|$ are $<10^{-6}$. These very small values indicate that there is very little 
interstellar polarisation toward this nearby star system (d=15pc). This is consistent with the result for 
55 Cnc and the results for the nearby stars given in Table 2 and in Hough et al.(2006).

Quasi-periodic millimagnitude photometric variations in the optical light from $\tau$ Boo have 
been detected by the MOST microsatellite (Walker et al.2008). This was attributed to strong star 
spot activity. Reduced flux due a large spot was seen in 2004, located
65$^{\circ}$ in advance of the sub-planetary point. In 2005 the amplitude of the flux variations was 
much smaller and on one occasion the spot appeared bright rather than dark.
They suggest that this is due to a magnetic interaction between the star and the planet, see also 
Shkolnik et al.(2005). 

The greater stability of our data in 2005 is consistent with the greater photometric stability in that 
year, if the polarisation variability is linked to star spots. Spot activity can 
induce weak linear polarisation in two ways. One way is through the transverse Zeeman effect (e.g.
Kemp \& Wolstencroft 1973; Borra \& Vaughn 1976) but since that is a line effect and the global 
magnetic field of $\tau$ Boo is fairly weak (Catala et al.2007) this seems unlikely. The other 
possibility is that star spots are breaking the symmetry of the polarisation of the stellar disc.
Stellar discs have a centrosymmetric polarisation pattern due to light scattering (Leroy 2000), 
with maximum polarisation very close to the limb of the star. In a uniform, spherically symmetric star 
the net polarisation will be zero but star spots near the limb can produce a measurable polarisation 
by breaking the symmetry, see Carciofi \& Magalhaes (2005).

We might therefore expect to see low level polarisation with a 3.3 day period but with differences 
between our 2004 and 2005 measurements. While no periodic variations in polarisation are detected, 
the data were less stable in 2004 than 2005. However, the dataset is too small to permit us to distinguish 
variations on a timescale of days from variations between these two years.
This issue clearly complicates any attempts to detect $\tau$ Boo b in polarised light.

\onecolumn
\hspace{-6mm}{\bf Table 4.} Individual measurements for 55 Cnc. Telescope Polarisation and Instrumental 
Polarisation have been subtracted. Orbital angle $\phi$=180$^{\circ}$ corresponds to minimum illumination
of the observed hemisphere.
\begin{center}
    \begin{tabular}{lcccccc}
  \hline
Date&  $\phi_Q$ &     Q/I    &  $\phi_U$    &       U/I   &  $<\phi>$ &    P \\
	&	& ($\times$10$^{-6}$)	 &       & ($\times$10$^{-6}$) &	     &	\\ \hline
15/2/06 & 264.8   &  -2.0$\pm$4.2  &  267.4  &	0.2$\pm$4.2    & 266.1   &	0.0$^{+4.2}$ \\
15/2/06 & 270.1   &  2.5$\pm$4.5   &  272.6  &	-8.7$\pm$4.6   & 271.4   &	7.8$\pm$4.6 \\
15/2/06 & 278.5   &  -2.1$\pm$4.5  &  281.1  &	8.5$\pm$4.4    & 279.8   &	7.5$\pm$4.5 \\
16/2/06 &  16.8   &  -2.0$\pm$4.2  &   19.3  &	-2.5$\pm$4.3   &  18.1   &	0.0$^{+4.3}$ \\
16/2/06 &  30.8  &  -6.8$\pm$4.2  &    33.4  &	-5.1$\pm$4.1   &  32.1  &	7.4$\pm$4.2 \\
16/2/06 &  36.0  &  0.1$\pm$4.5   &    38.6  &	-4.7$\pm$4.2   &  37.3  &	1.6$\pm$4.4 \\
16/2/06 &  41.1  &  -4.3$\pm$4.3  &    43.7  &	-5.3$\pm$4.5   &  42.4  &	5.2$\pm$4.4 \\
17/2/06 & 140.9   &   -1.3$\pm$4.1 &   143.4 &	0.5$\pm$4.0    & 142.2  &	0.0$^{+4.1}$ \\
17/2/06 & 146.0   &   -4.2$\pm$4.3 &   148.5 &	-4.1$\pm$4.3   & 147.2   &	4.0$\pm$4.3 \\
17/2/06 & 157.1   &   4.0$\pm$4.5  &   159.7 &	-3.2$\pm$4.6   & 158.4   &	2.4$\pm$4.6 \\
17/2/06 & 162.2   &   2.2$\pm$4.3  &   164.8 &	-1.0$\pm$4.1   & 163.5   &	0.0$^{+4.2}$ \\
17/2/06 & 167.4   &   -5.5$\pm$4.6 &   169.9 &	-7.6$\pm$4.3   & 168.7   &	8.2$\pm$4.5 \\
18/2/06 & 269.6   &  1.0$\pm$4.1   &   272.2 &	0.2$\pm$4.5    & 270.9   &	0.0$^{+4.3}$ \\
18/2/06 & 282.5   &  6.4$\pm$4.3   &   285.0 &	-0.8$\pm$4.6   & 283.8   &	4.7$\pm$4.5 \\
18/2/06 & 287.5   &  3.1$\pm$4.3   &   290.1 &	-10.7$\pm$4.3  & 288.8   &	10.3$\pm$4.3 \\
18/2/06 & 292.8   &  2.8$\pm$4.9   &   295.3 &	-12.3$\pm$4.8  & 294.1   &	11.7$\pm$4.8 \\
18/2/06 & 297.9   &  0.1$\pm$5.3   &   300.5 &	-11.0$\pm$5.1  & 299.2   &	9.7$\pm$5.2 \\
19/2/06 &  39.6  &  -1.2$\pm$4.5  &     42.1 &	-7.4$\pm$4.2   &  40.9   &	6.1$\pm$4.4 \\
19/2/06 &  53.9  &  3.3$\pm$4.2   &     56.5 &	-0.2$\pm$5.3   &  55.2  &	0.0$^{+4.8}$ \\
19/2/06 &  65.6  &  3.0$\pm$4.4   &     68.2 &	-7.6$\pm$4.1   &  66.9  &	6.9$\pm$4.2 \\
20/2/06 & 160.0   &  5.5$\pm$4.1   &   162.6 &	1.9$\pm$4.3    & 161.3  &	4.0$\pm$4.2 \\
20/2/06 & 165.2   &  -2.7$\pm$4.2  &   167.8 &	-8.8$\pm$4.2   & 166.5   &	8.2$\pm$4.2 \\
20/2/06 & 171.4   &  -6.6$\pm$4.3  &   176.6 &	-0.5$\pm$3.2   & 174.0   &	5.4$\pm$3.8 \\
20/2/06 & 179.9   &  9.4$\pm$4.1   &   182.4 &	-8.9$\pm$4.3   & 181.2   &	12.2$\pm$4.2 \\
20/2/06 & 194.1   &  -0.6$\pm$4.4  &   196.7 &	-2.4$\pm$4.3   & 195.4   &	0.0$^{+4.4}$ \\ \hline
\end{tabular}
\end{center}
\twocolumn

\onecolumn
\hspace{-6mm}{\bf Table 5.} Individual measurements for $\tau$ Boo. Telescope Polarisation and 
Instrumental Polarisation have been subtracted. Orbital angle $\phi$=180$^{\circ}$ corresponds to minimum 
illumination of the observed hemisphere.
\begin{center}
    \begin{tabular}{lcccccc}
  \hline
Date   &  $\phi_Q$  & Q/I    &         $\phi_U$ &    U/I    &     $<\phi>$ &  P \\
	&	& ($\times$10$^{-6}$)	 &       & ($\times$10$^{-6}$) &	     &	\\ \hline
24/04/04  &  -22.47 & -4.9 $\pm$2.5	&   -25.17 & -0.1 $\pm$2.6 &	-23.82 &	4.6$\pm$2.7 \\
24/04/04  &  -17.82 & 2.5 $\pm$2.2	&  -20.0  & -8.4 $\pm$2.3 &	-18.91 &	8.2$\pm$2.5 \\
24/04/04  &  -9.09  & 6.0 $\pm$2.6	&   -6.71 & -1.2 $\pm$2.3 &	-7.9   &	5.8$\pm$2.5 \\
25/04/04  &  75.69  & 3.6 $\pm$2.5      &   77.99 & -0.7 $\pm$2.3 &	76.84  &	2.9$\pm$2.5 \\
25/04/04  &  80.44  & 3.9 $\pm$2.7	&   83.82 &  3.1 $\pm$2.5 &	82.13  &	4.7$\pm$2.5 \\
25/04/04  &  89.97  & 0.0 $\pm$2.4	&   92.13 &  0.8 $\pm$2.3 &	91.05  &	0.0$^{+2.5}$ \\
25/04/04  &  100.7  & 11.1 $\pm$2.4 	&   97.96 &  0.7 $\pm$2.3 &	99.35  &	11.0$\pm$2.5 \\
26/04/04  &  -165.1 & 3.0 $\pm$2.6 	& -162.8  & -17.1 $\pm$2.6 &	-163.9 &	17.5$\pm$2.5 \\
26/04/04  &  -160.5 & 1.9 $\pm$2.5     & -158.2  & -16.2 $\pm$2.5 &	-159.4 &	15.8$\pm$2.5 \\
26/04/04  &  -149.2 & 14.1 $\pm$2.6    & -151.5  & -7.6 $\pm$2.4 &	-150.4 &	15.8$\pm$2.5 \\
25/04/05  &  -103.3 & -2.3 $\pm$2.5    & -101.1  & -1.6 $\pm$2.5 &	-102.2 & 0.0$^{+2.5}$ \\
26/04/05  &  6.5    & 3.8 $\pm$3.9     & 10.3  &  13.8 $\pm$2.8 &	8.4    & 14.1$\pm$3.4 \\
26/04/05  &  18.5   & 3.1 $\pm$2.4    &	20.5  &  4.6 $\pm$2.5 	&	19.5  & 5.2$\pm$ 2.5 \\
26/04/05  &  23.0   & 2.5 $\pm$2.5    &	25.1  &  4.7 $\pm$2.4 	&	24.05 & 4.9$\pm$2.5 \\
27/04/05  &  121.7  & -4.9 $\pm$2.4   &	124.0 &  3.2 $\pm$2.5 	&	122.8 & 5.3$\pm$2.5 \\
27/04/05  &  126.1  & -1.7 $\pm$2.9   &	128.1 &  1.9 $\pm$2.5 	&	127.1 & 1.3$\pm$2.7 \\
28/04/05  &  -146.6 & -0.7 $\pm$2.6   &	-143.9 & -2.3 $\pm$2.5 	& 	-145.2 & 0.8$\pm$2.6 \\
28/04/05  &  -137.0 & -2.0 $\pm$2.4   &	-134.5 & 3.9 $\pm$2.5 	&	-135.8 & 3.9$\pm$2.5 \\
28/04/05  &  -127.3 & -3.8 $\pm$2.5   &	-125.0 & 8.6 $\pm$2.4 	&	-126.1 & 9.0$\pm$2.5 \\
29/04/05  &  -39.2  & -0.0 $\pm$2.5   &	-37.2 &  0.4 $\pm$2.5 	&	-38.2  & 0.0$^{+2.5}$ \\
29/04/05  &  -26.3  & -2.0 $\pm$2.5   &	-24.0 &  3.5 $\pm$2.6 	&	-25.15 & 3.4$\pm$2.5 \\
29/04/05  &  -22.4  & -1.8 $\pm$2.5   &	-19.7 &  -1.1 $\pm$2.5 	& 	-21.05 & 0.0$\pm$2.5 \\
29/04/05  &  -14.7  & -3.5 $\pm$2.4   &	-12.7 &  5.7 $\pm$2.5 	&	-13.7  & 6.3$\pm$2.5 \\
30/04/05  &  69.7   & 0.8 $\pm$2.6    &	67.5  &  0.5 $\pm$2.6 	&	68.6  & 0.0$^{+2.6}$ \\
 7/05/05  &  116.6  & -1.9 $\pm$3.5   &	118.6 &  3.2 $\pm$3.5 	&	117.6 & 2.9$\pm$3.5 \\
 7/05/05  &  120.9  & 3.8 $\pm$4.0   &	123.0 &  4.5 $\pm$3.5 	&	121.9 & 5.4$\pm$3.8 \\
 7/05/05  &  127.5  & 4.1 $\pm$3.7   &	125.4 &  4.3 $\pm$3.5 	&	126.4 & 5.4$\pm$3.7 \\
 8/05/05  &  -141.9 & -4.4 $\pm$2.6  &	-139.5 & 4.4 $\pm$2.5 	&	-140.7 & 6.0$\pm$2.6 \\
 8/05/05  &  -137.5 & -5.7 $\pm$2.3  &	-135.4 & 11.2 $\pm$2.7 	&	-136.4 & 12.6$\pm$2.5 \\
 8/05/05  &  -127.0 & -0.9 $\pm$2.5  &	-125.0 & 4.0 $\pm$2.5 	&	-126.0 & 3.2$\pm$2.5 \\
 8/05/05  &  -122.7 & -6.0 $\pm$2.4  &	-120.5 & 8.6 $\pm$2.5 	&	-121.6 & 10.3$\pm$2.5 \\
 8/05/05  &  -118.4 & -1.7 $\pm$2.7  &	-116.3 & 9.8 $\pm$2.5 	&	-117.4 & 9.8$\pm$2.6 \\
16/02/06  &  55.3   & 9.1 $\pm$2.6   &	57.5   & -12.1$\pm$2.3 	& 	56.4  & 15.0$\pm$2.5 \\
17/02/06  &  159.4  & -1.2 $\pm$2.6  &	161.6  & 2.1 $\pm$2.5 	&	160.5 & 0.0$^{+2.6}$ \\
18/02/06  &  -88.2  & 6.2 $\pm$2.5   &	-86.0  & -4.3 $\pm$2.6 	& 	-87.1 & 7.2$\pm$2.6 \\
19/02/06  &  15.3   & 6.4 $\pm$2.6   &	17.5  &  4.7 $\pm$2.7 	&	16.4  & 7.5$\pm$2.7 \\
20/02/06  &  123.7  & 3.5 $\pm$2.5   &	125.9  &  7.5 $\pm$2.5 	& 	124.8 & 7.8$\pm$2.5 \\
21/02/06  &  -122.0 & 4.8 $\pm$2.5   &	-119.8 & -1.7 $\pm$2.5 	& 	-120.9 & 4.5$\pm$2.5 \\ \hline
\end{tabular}
\end{center}
\twocolumn

\section{Modelling and Discussion}

\subsection{Initial assumptions}

We model the possible polarisation signatures of extrasolar planets with a Monte
Carlo multiple scattering code with a radiative transfer similar to that used in 
Seager, Whitney \& Sasselov (2000) and described in Green et al.(2003). The finite 
angular size of the star is included, with a solar limb darkening model (Claret 2000), 
and the eccentricity of the orbit is set to zero. All photons are assumed to hit the 
star-facing side of the planetary disc at randomly generated locations. Unlike Seager et al. 
we do not use a full model atmosphere with precisely specified composition and pressure-temperature 
profile. These quantities are hard to estimate and many of the variables will not affect 
the broad band polarised light signal of the planet. Instead we simply assume that the 
broad band radiative transfer will be dominated by Rayleigh scattering by
molecules or Mie scattering by dust particles and we do not explicitly include molecular or atomic 
absorption. Our code assumes a uniform, locally plane parallel semi-infinite atmosphere 
with no variation of density with depth. In such a geometry, adding molecular or atomic 
absorption features would be equivalent to reducing the single scattering albedo.

\begin{figure}
\vspace{-7.2cm}\includegraphics[scale=1.0,angle=0]{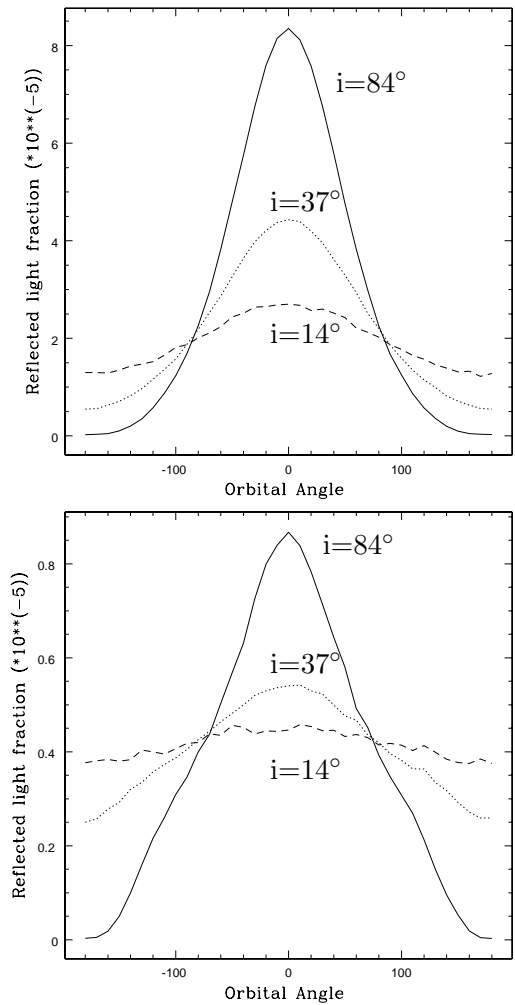}
\vspace{-9.5cm}
\caption{Reflected light curves. The reflected light from the planet is
divided by the light from the star to give the fractional change in flux
from the star$+$planet system. Solid lines: i=84$^{\circ}$. Dotted
lines: i=37$^{\circ}$. Dashed lines: i=14$^{\circ}$. {\it (upper panel)} 
``Rayleigh scattering'' model, using $\varpi$=0.99. {\it (lower panel)} 
Interstellar dust (ISD) model. The ISD model produces an order of magnitude
less reflected light due to weaker back scattering in the phase function and a lower
single scattering albedo, see text $\S$4.2. Note that the curves are not perfectly 
symmetric about $\phi$=0 due to shot noise in the Monte Carlo simulations.}
\end{figure}

\begin{figure}
\vspace{-7.2cm}\includegraphics[scale=1.0,angle=0]{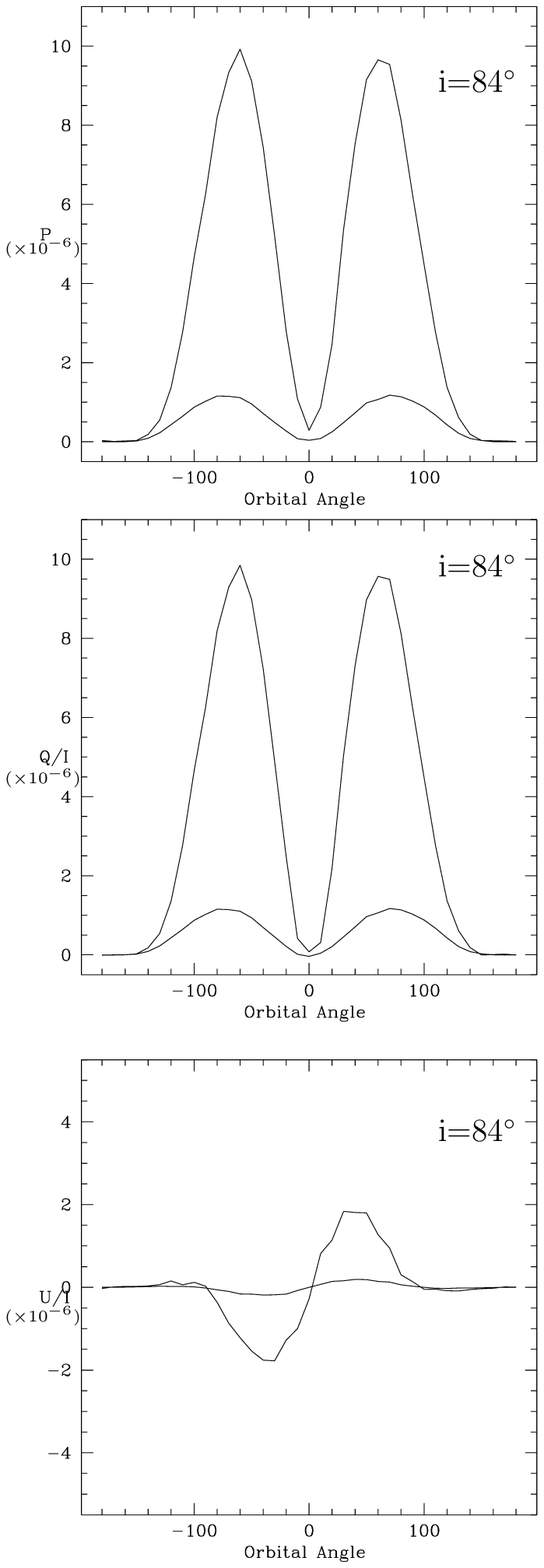}
\vspace{-2cm}
\caption{Model polarisation for orbital inclination $i=84^{\circ}$.
{\it (top)} fractional polarisation, P; {\it (middle)} Q/I; 
{\it (bottom)} U/I. The curves with larger amplitude are for
a Rayleigh scattering atmosphere. The curves with smaller amplitude
are for an interstellar dust model (see text).}
\end{figure}

\begin{figure}
\vspace{-7.2cm}\includegraphics[scale=1.0,angle=0]{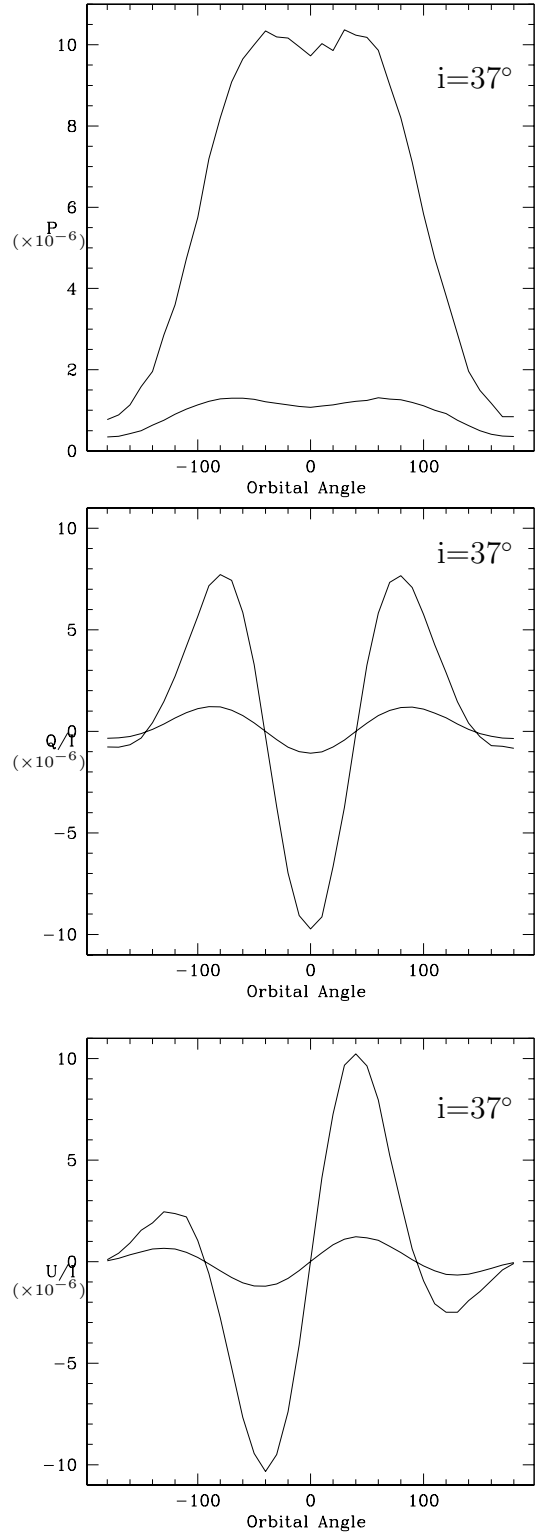}
\vspace{-2cm}
\caption{Model polarisation for orbital inclination $i=37^{\circ}$.
{\it (top)} fractional polarisation, P; {\it (middle)} Q/I; 
{\it (bottom)} U/I. The curves with larger amplitude are for
a Rayleigh scattering atmosphere. The curves with smaller amplitude
are for an interstellar dust model (see text). Note that the plots
of P and Q/I are not perfectly symmetric about $\phi$=0 due to shot
noise in the Monte Carlo simulations.}
\end{figure}

\begin{figure}
\vspace{-7.2cm}\includegraphics[scale=1.0,angle=0]{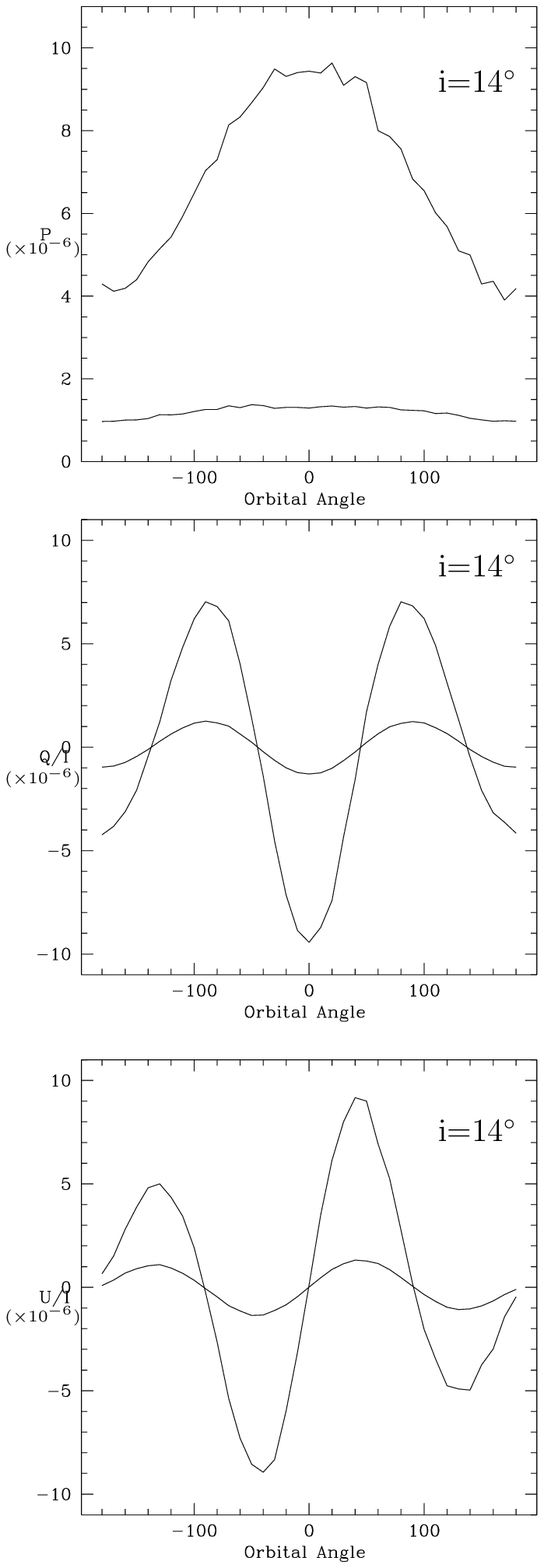}
\vspace{-2cm}
\caption{Model polarisation for orbital inclination $i=14^{\circ}$.
{\it (top)} fractional polarisation, P; {\it (middle)} Q/I; 
{\it (bottom)} U/I. The curves with larger amplitude are for
a Rayleigh scattering atmosphere. The curves with smaller amplitude
are for an interstellar dust model (see text).}
\end{figure}

In real atmospheres, molecular and atomic absorption features are pressure sensitive,
whereas the single scattering albedo of dust particles is not, so their effects cannot
simply be combined in a single parameter for each wavelength (see Prather (1974) and references 
therein). However, our assumption of a uniform atmosphere allows us to express in a simple manner
the relation between the data and attributes of planets such as the geometric albedo and the radius.

For the individual scattering events, the models considered here have either exactly the Rayleigh scattering 
phase functions and Rayleigh angular dependence of polarisation or else are Rayleigh-like. 
The exact Rayleigh phase functions and polarisation behaviour is appropriate for either Rayleigh 
scattering by molecules or scattering by spherical dust particles (Mie scattering) that are much smaller than 
the wavelength of observation. This is because Mie scattering becomes identical to Rayleigh scattering
in the limit of small particle size, except in one respect, which is that the albedo of very small
dust particles will usually be less than unity. The {\it Rayleigh-like} models are Mie scattering models
in which the size of the optically dominant particles is smaller than the wavelength but not small
enough to be in the Rayleigh limit. In this size regime Mie scattering has a similar phase function
to Rayleigh scattering but a higher probability to scatter photons forward rather than backward.
The angular dependence of polarisation is also similar to Rayleigh scattering (a bell-shaped curve
with maximum polarisation occuring at scattering angles near 90$^{\circ}$) but the maximum polarisation
is less than 100\%.

In this regime, the most important model atmosphere parameters governing the polarised flux reflected 
from the planet are the single scattering albedo, $\varpi$ and the maximum fractional polarisation produced
by scattering of unpolarised light through 90$^{\circ}$, $p_m$. As noted above, reducing $\varpi$ is equivalent
to introducing unspecified molecular absorption into the atmosphere. Models with larger dust grains
produce a very different behaviour, generally with a lower polarised flux (see Seager et al.2000). 
However, such models are somewhat less probable because any large grains
are predicted to settle below the photosphere under gravity (e.g. Helling, Woitke \& Thi 2008; 
Ackerman \& Marley 2001). The atmospheres of Jupiter and Saturn exhibit a combination of Rayleigh 
scattering and scattering by aerosols, some of which are small compared to the $\sim$0.78~$\mu$m wavelength 
of the filter employed here and some of which are larger, e.g. Leroy (2000) and references therein.

The orbital radius, $r$, is an additional parameter in our code. For very small orbital radii the
incident radiation field becomes significantly non-parallel. However, we find that for the orbital
radii of the planets considered in this paper ($r$$\ge$0.038~AU) this makes a negligible difference
to the results, by comparison with the plane parallel case for the same orbital radius. This extends
the result of Seager et al.(2000), who also found this to be the case for the reflected flux curves
when dealing with reasonably isotropic phase functions but did not comment on the effect on the
polarised light curves. However, Seager et al.(2000) and Green et al.(2003) found that a non-parallel
radiation field does lead to significantly different reflected light curves for models that use 
highly forward throwing phase functions.

We have bench tested our code against the results of Prather (1974) and Stam et al.(2006). Prather (1974)
calculated geometric albedo as a function of $\varpi$ for a planet with a semi-infinite Rayleigh 
scattering atmosphere, assuming a plane parallel incident radiation field. Stam et al.(2006) used a
locally analytical method combined with a numerical method for integrating over the planetary disc 
to calculate both polarisation and the reflected light curve for a planet with a 
pure Rayleigh scattering atmosphere ($\varpi$=1) and an optical depth of 5.75, with a black surface below. 
Again, a plane parallel incident radiation field was assumed. Our code accurately reproduces the 
polarisation curve shown in Figure 7 of Stam et al.(2006), including the polarisation reversal that
occurs at small scattering angles (i.e. close to the ``new moon'' phase) due to multiple scattering
effects. Our calculated geometric albedos agree with those of Prather (1974) to within 5\%. In addition,
our fractional polarisation curves for Rayleigh scattering models appear to be in good agreement
with those shown in Seager et al.(2000) for atmospheres with 0.1~$\mu$m grains, although a precise
quantitative comparison is not possible because of the much more complicated model atmospheres used in that
work. The method of Stam et al.(2006) has the advantage that it can produce results with excellent precison
for homogeneous atmospheres, whereas the Monte Carlo method has low signal to noise for calculations
close to the new moon phase of a planet, owing to the small number of photons. However, our
Monte Carlo approach has the advantages of (i) greater flexibility to treat vertically and horizontally 
inhomogeneous atmospheres and (ii) the ability to treat planets very close to the central star, where the
incident radiation field is not plane parallel.

\subsection{Sensitivity to planets}

In Figures 5, 6, 7 and 8 we show the reflected light curves and the fractional P, Q/I and U/I outputs from 
two 20 million photon runs with different model atmospheres. Both simulations are for a planet with radius 
$R$=1.2~R$_{Jup}$ (which is the typical size of a hot Jupiter) in a circular orbit at $r$=0.05~AU around a 
star with radius $R_{*}$=7$\times$10$^8$m. In both cases the positive Stokes Q plane lies parallel
to the axis of the orbit. Figure 6 shows the polarisation when the system has an orbital inclination 
$i=84^{\circ}$,  Figure 7 shows it for $i=37^{\circ}$ and Figure 8 shows it for $i=14^{\circ}$. The upper panel
of Figure 5 and the curves with larger amplitude in Figures 6-8 are for a run approximating a pure Rayleigh scattering 
atmosphere, the only difference being that the adopted single scattering albedo is $\varpi$=0.99 rather than unity,
in order to ensure that the run completed in a finite time. The lower panel of Figure 5 and the curves with 
smaller amplitude in Figures 6-8 are for an atmosphere composed entirely of interstellar dust grains, using a
silicate and graphite interstellar dust model taken from Fischer, Henning \& Yorke (1994) in which the 
grains have sizes between 0.005 and 0.25~$\mu$m and an $n(r)\propto r^{-3.5}$ distribution within that range. 
This model is run for a wavelength of 0.78~$\mu$m, at which $\varpi$=0.57 and $p_m$=0.4. 

Figure 6(a) shows that maximum polarisation of the planetary system occurs near orbital angle 
$\phi$=$\pm 60^{\circ}$ (a gibbous phase) for orbital inclinations close to 90$^{\circ}$. This is because the 
fractional polarisation of light reflected by the planet (which is greatest at $\phi$=$\pm 90^{\circ}$)
is modulated by the reflected light curves in Figure 5 when calculating the fractional polarisation of the 
star$+$planet system. The maximum polarised flux from the planet therefore occurs at a gibbous phase. 

The models show that Rayleigh scattering and Mie scattering by small grains produce
polarisation signals with a qualitatively similar dependence on orbital angle, $\phi$. The maximum
fractional polarisation, P, produced by the interstellar dust model is approximately a tenth of that 
produced by the Rayleigh scattering model. This is due to the combination of the smaller values of $\varpi$ 
and $p_m$ and weaker backscattering in the phase function. The interstellar dust mixture has a
moderately forward throwing phase function at 0.78~$\mu$m, described by the asymmetry parameter
$g = $$<$$cos(\theta_S)$$>$ = 0.38, where $\theta_S$ is the scattering angle. This compares with $g$=0 for 
Rayleigh scattering, which has the phase function P($\theta_S$)= 3 (1 $+ cos^2(\theta_S)$)$/16 \pi$, where
P($\theta_S$) is the scattering probability per unit solid angle.

Prather (1974) and Sromovsky (2005) have shown that increasing the single scattering albedo has a highly 
non-linear effect on the total flux reflected by a planet at the full moon phase (defined here as 
$\phi$=0$^{\circ}$). The 
reflected flux (polarised flux plus unpolarised flux) at this phase is described by the geometric 
albedo, $A_G$, which is equal to unity for a flat isotropically scattering disc (a Lambert disc).
For a pure Rayleigh scattering atmosphere $A_G$=0.80, whereas for our Rayleigh-like model
with $\varpi$=0.99 the expected value declines to $A_G$=0.64 (Prather 1974). This is because
a significant fraction of photons incident upon the planet require a large number of scattering 
events to escape the atmosphere. When $\varpi$$<$0.5, $A_G$ is closer to a linear function of single 
scattering albedo, but at larger values of $\varpi$ multiple scattering causes $A_G$ to increase rapidly.

Although multiple scattering can greatly increase the geometric albedo, it also has a depolarising
effect. This is because multiply scattered photons usually have lower polarisations than singly
scattered photons, and the plane of polarisation typically differs also. 

In order to determine how to relate our polarisation measurements to geometric albedos we have 
investigated the dependence of the polarisation on $\varpi$ and the orientation of the orbit in space.
To derive meaningful upper limits from the data in $\S$3 we must consider 
not the absolute value of P=$\sqrt{(Q^2+U^2)}/I$ but the expected changes in Q/I and U/I during an 
orbital period, and the average of these over the range of possible orbital inclinations and orientations of
the orbit projected in the plane of the sky.

\begin{figure}
\hspace{-7.5cm}\includegraphics[scale=1.0,angle=0]{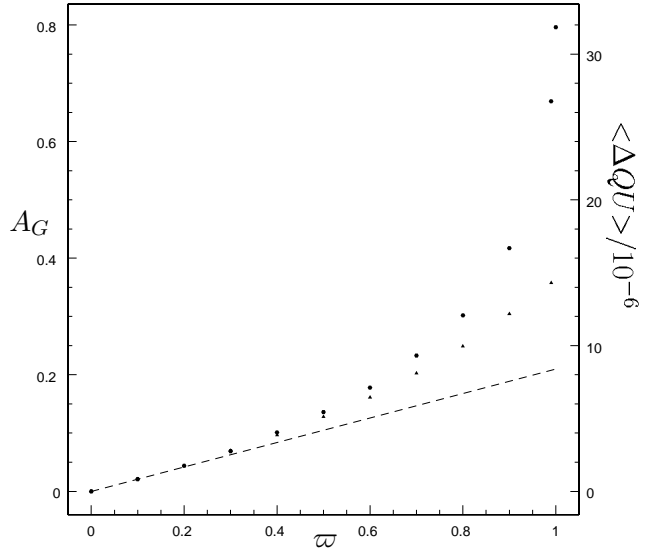}
\vspace{-15cm}
\caption{Geometric albedo, $A_G$ and expected peak to trough polarisation modulation,
$<\Delta QU>$, for a Rayleigh scattering atmosphere as a function of single scattering
albedo, $\varpi$. The upper points (circles) represent $A_G$, shown on the left axis. The 
lower points (triangles) represent $<\Delta QU>$, shown on the right axis.}
\end{figure}

\begin{table}
\setcounter{table}{5}
\caption{Predicted geometric albedo, $A_G$, and expected polarisation properties as a 
function of single scattering albedo for planets with Rayleigh-like atmospheres (see text). The polarisation 
properties are calculated for a planet with $R$=1.2~$R_{Jup}$ and $r$=0.05~AU and are given in units
of 10$^{-6}$.}
\begin{center}
\begin{tabular}{cccc}
  \hline
$\varpi$ & $A_G$ & $<\Delta QU>/10^{-6}$ & $<$$P_{MV}$$>/10^{-6}$ \\ \hline
1.0	& 0.796$^a$	& - & -	\\
0.99	& 0.669 & 14.30	& 10.10 \\
0.9	& 0.417 & 12.17	&  8.40 \\
0.8	& 0.302 & 9.95	&  6.95 \\
0.7	& 0.233 & 8.10	&  5.63 \\
0.6	& 0.178 & 6.45	&  4.55 \\
0.5	& 0.136 & 5.12	&  3.57 \\
0.4	& 0.101 & 3.87	&  2.69 \\
0.3	& 0.069 & 2.79	&  1.93 \\
0.2	& 0.044 & 1.75	&  1.22 \\
0.1	& 0.021 & 0.84	&  0.59 \\ \hline
\end{tabular}
\end{center}
\small
Note: (a) The geometric albedo of 0.796 for a pure Rayleigh scattering atmosphere with $\varpi$=1.0 is taken 
from Prather (1974).
\end{table}

We define the quantity $<$$\Delta QU$$>$ as the peak to trough polarisation modulation that we would 
typically expect to observe in at least one of Stokes Q/I or U/I for a planet with a Rayleigh scattering
phase function and angular dependence of polarisation, after averaging over the ensemble of possible
orbital inclinations and orientations in proportion to their probability (Prob$(i)$$\propto$$sin(i)$).
The maximum value of P during an orbit, denoted P$_{MV}$, is almost independent of orbital inclination, 
as shown by Seager et al.(2000). P$_{MV}$ increases very slightly with decreasing orbital inclination,
e.g. it is 7\% higher at i=29$^{\circ}$ ($cos(i)=0.9$) than at i=87$^{\circ}$ ($cos(i)$=0.05).
At large inclinations ($cos(i) \le 0.25$) the polarisation lies mainly 
in one plane (see Figure 6(a-c)) and has a maximum of
P$_{MV}$$\approx$1.2$ \times 10^{-6}(\varpi/0.2)(r/0.05 AU)^{-2} (R/1.2R_{Jup})^2$, for small values of 
$\varpi$, where $r$ is orbital radius and $R$ is the planetary radius.
This would be detected as a maximum positive or negative excursion in Q/I or U/I from the 
small constant interstellar polarisation, which would occur near $\phi$=$\pm$60$^{\circ}$.
Considering the range of possible orientations of the orbital as projected in the plane of
sky, either Q/I and/or U/I would have a variation of at least P$_{MV}/\sqrt{2}$,
each having a statistical average of 0.9~P$_{MV}$. At smaller inclinations (see Figures 7 and 8)
we see larger variations: both Q/I and U/I have variations $\ga$ 1.5~P$_{MV}$ over the course 
of an orbit as they change periodically from positive to negative in sign. For inclinations $cos(i) \ge$0.25 
we take the average of the full variations in Q/I, U/I and in the intermediate polarisation planes
at position angles of 22.5$^{\circ}$ and 67.5$^{\circ}$ to predict the amount that would
be observed. The magnitude of the variations in all these possible planes of polariation is similar. 
In the limiting case of $i$=0 we would expect that P is constant, while Q/I and U/I have equal amplitude, 
since the planet would always show a half moon phase in this 'pole-on' view of the orbit.

Averaging over the full range of inclinations, (weighting by their probability), we find that the 
typical peak to trough variation that we would expect to observe is 
$<\Delta QU>$$\approx$1.75$ \times 10^{-6}(\varpi/0.2)(r/0.05 AU)^{-2} (R/1.2R_{Jup})^2$, for small
values of $\varpi$. Since the orbital inclination has only a modest effect on the sensitivity of 
polarisation measurements to extrasolar planets (when searching for periodic variations in the Stokes 
parameters) it is acceptable to use this value when placing limits on planets whose orbital inclination.
is unknown.

In Table 6 and Figure 9 we show how $<$$\Delta QU$$>$, $A_G$ and $<$$P_{MV}$$>$ change as functions of 
$\varpi$, for a planet at $r$=0.05~AU with radius $R$=1.2$R_{Jup}$. ($<$$P_{MV}$$>$ is the value of 
P$_{MV}$ obtained after averaging over the range of orbital inclinations, i.e. it is the typical 
maximum polarisation of the planetary system that we would expect to observe in the course of an orbit, 
in the absence of polarisation due to the interstellar medium.) All these variables are close to 
linear functions of $\varpi$ when $\varpi$ is small, but increase
more rapidly as $\varpi$ approaches unity owing to the growing contribution of multiply scattered
photons to the reflected flux. We find that $<$$\Delta QU$$>$ and $<$$P_{MV}$$>$ are closer than $A_G$ to 
being linear functions of $\varpi$, owing to the depolarising effect of multiple scattering at high values
of $\varpi$, which partly counterbalances the increased reflected flux. $<$$\Delta QU$$>$ and $<$$P_{MV}$$>$ 
have essentially identical dependences on $\varpi$ and obey the simple relation
$<$$\Delta QU$$>$=1.43$<$$P_{MV}$$>$ to within the $\sim$1\% uncertainty imposed by the these Monte Carlo 
simulations with 2$\times$10$^7$ photons. Wood et al.(1996) found
a qualitatively similar dependence of polarisation on $\varpi$ when modelling the scattering of light
in a geometrically thin circumstellar accretion disc. The behaviour held true for all inclination angles
of the disc. 

The relation between $<$$\Delta QU$$>$ (the typical peak to trough polarisation modulation) and 
$A_G$ is almost linear for $\varpi$$\le$0.7. The data in Table 6 show that the relation 
$<$$\Delta QU$$>$=3.9$\times$$10^{-5}$$A_G$ is valid to within $\sim$10\% for $\varpi$$<$0.7, or $A_G$$<$0.22. 
For the more general case where $R$ and $r$ may be different and scattering may be only Rayleigh-like rather
than Rayleigh (see $\S$4.1), this becomes\\

\hspace{-6mm}(1) \hspace{2mm} $<$$\Delta QU$$>$=3.9$\times$$10^{-5}$$p_m A_G (r/0.05 AU)^{-2}(R/1.2R_{Jup})^2$\\

Table 6 shows that when $\varpi$=0.99, P$_{MV}$$\approx$1.0$\times 10^{-5}$ for a planet at $r$=0.05~AU
with $R$=1.2$R_{Jup}$. For a transiting planet such as HD189733~b with $i$$\approx$90$^{\circ}$ the peak 
to trough variation in Q/I and U/I would typically be 0.9~P$_{MV}$, as noted earlier, and the maximum 
possible value would be P$_{MV}$. HD189733~b has $r$=0.031~AU and $R$=1.2~$R_{Jup}$ (Miller-Ricci et 
al.2008; Winn et al.2007; Pont et al.2007; Baines et al.2007) so when we scale to the smaller orbital
radius we find that the expected peak to trough variation would typically be 2.3$\times 10^{-5}$, 
with a maximum possible value of 2.6$\times10^{-5}$ for this Rayleigh scattering model with $\varpi$=0.99.
It is therefore hard to see how to explain the variations in Q/I and U/I at a level $\ge$10$^{-4}$ that
were reported by Berdyugina et al.(2008).

\normalsize
\subsection{Upper limits on the planets around 55 Cnc and $\tau$ Boo}

The inclination and orientation of the projected orbits of the planets around 55 Cnc is not known with 
high precision or confidence so we cannot predict whether a variation should more easily be seen in Q/I or U/I.
The inclination given in Table 1 is only an estimate. By contrast, the inclination of the orbit
of $\tau$ Boo~b is constrained to be $i$$\sim$40$^{\circ}$ with fair confidence (see $\S$1).

Inspection of Figures 2 and 3 shows that the datasets have far from complete phase coverage.
Nonetheless, the datasets for both 55 Cnc e and $\tau$ Boo b are sensitive to the full variations in 
Stokes Q/I and/or U/I predicted for all possible inclinations. For 55 Cnc b, which has a longer orbital
period (see Table 1), the range of orbital angles covered by the data is 206$^{\circ}$ to 334$^{\circ}$.
This range of angles samples between 67\% and 100\% of the predicted peak to trough variation in at least one
of Q/I and U/I, depending on the orbital inclination. The average value is approximately 75\% and we use this
value to derive the upper limit for that planet.

No variations of the nightly averages are detected that are consistent with those that might be 
expected from Figures 6, 7 and 8. Figure 7(a-c) illustrates the likely form of the
variations for $\tau$ Boo. However, it should be noted that the orientation of the axis of the projected 
orbit in the plane of the sky need not lie parallel to the Stokes Q or U planes. If it lies at a 
position angle of 22.5$^{\circ}$ (compared to 0$^{\circ}$ for Q and 45$^{\circ}$
for U) then the expected signal has the form of the average of the Q/I and U/I curves, multiplied
by $\sqrt{2}$.

We use the standard deviation of the nightly averages to determine 
an upper limit on the polarisation variations in both planets, including only nights with at least two 
data points. The standard deviations of the Q/I and U/I nightly averages for 55 Cnc are 
$\sigma_{n-1}=2.16\times 10^{-6}$ and $\sigma_{n-1}=2.31\times 10^{-6}$ respectively. Taking the 
average of these gives us a standard deviation of $2.24\times 10^{-6}$. 
For the hot Neptune 55 Cnc e this corresponds to a 4$\sigma$ sensitivity limit of 
$A_G$$<$0.13$(R/1.2 R_{Jup})^{-2}$$p_m^{-1}$, using eq.(1) from $\S$4.2, which is valid for 
planets with low albedos.
If we assume a somewhat smaller radius, e.g. 0.8$R_{Jup}$, for 55 Cnc e then the 4$\sigma$ upper limit
corresponds to a larger single scattering albedo and the formula is no longer valid.
E.g. for the case $R$=0.8$R_{Jup}$ the 4$\sigma$ limit is $A_G$$\la 0.38 p_m^{-1}$. If
the radius is even smaller then the data no longer place strong limits on $A_G$.
 
For the hot Jupiter 55 Cnc b the data do not provide a useful upper limit. For a highly reflective planet
at $r$=0.05~AU with $\varpi$=0.99, Table 6 indicates that $<$$\Delta QU$$>$=1.43$\times$$10^{-5}$.
After scaling this to the actual orbital radius of 0.115 AU and correcting for the 75\% factor caused by
the limited phase coverage, the expected peak to trough polarisation variation is only
$<\Delta QU>$=2.0$\times$$10^{-6}$, which is slightly less than the 1$\sigma$ sensitivity of the data.

For $\tau$ Boo the standard deviations of the Q/I and U/I nightly averages are 
$\sigma_{n-1}=3.62\times 10^{-6}$ and $\sigma_{n-1}=6.49\times 10^{-6}$ respectively. Taking the 
average of these gives us a standard deviation of $5.06\times 10^{-6}$. Assuming that the orbital 
inclination is $i$$\sim$40$^{\circ}$, the expected peak to trough variation in the Q/I and U/I parameters
is 5.1$\times$$10^{-5}$$A_G$ (taking the average of the variations in Q/I, U/I and the polarisation
planes at 22.5$^{\circ}$ and 67.5$^{\circ}$).
This leads to a 4$\sigma$ sensitivity limit of $A_G$$<$0.37$(R/1.2 R_{Jup})^{-2}$$p_m^{-1}$.
We note that the smaller
standard deviations for subset of data taken in 2005 for this planet could in principle be used to
establish a more sensitive limit. However, the phase coverage in the 2005 data is not fully sensitive to
the predicted polarisation variations so we adopt the larger standard deviations
associated with the full dataset.

The upper limit for $\tau$ Boo b is similar to those previously reported at shorter visible wavelengths
by Charbonneau et al.(1999) and Leigh et al.(2003), based on searches for Doppler shifts in stellar 
absorption lines after reflection by the planet.

These datasets therefore provide useful upper limits on the reflected light from  55 Cnc e and
$\tau$ Boo b but not for 55 Cnc b, which has a larger orbit. It is notable that the data for
$\tau$ Boo are more than a factor of two less stable than the data for 55 Cnc, even though the $\tau$ 
Boo data points have 42\% smaller uncertainties due the smaller photon noise for the brighter star. 
The scatter in the 55 Cnc data is consistent with the internal errors, whereas the scatter in the
$\tau$ Boo data is not, see $\S$3.2.

\subsection{Stability of interstellar polarisation}

The relatively large dataset on bright nearby stars taken in February 2006 provides
some redundancy in the fit to telescope polarisation. Assuming the sample of five stars
in Table 3 is representative, we can infer that nearby stars with a wide range of spectral types 
have no variations from night to night above a typical 1$\sigma$ level of 
$\sim 2 \times 10^{-6}$. In calculating this figure we have excluded from the nightly
averages ten outlying measurements (13\% of the data) that depart from the TP fit 
by $>$3$\sigma$ and $>$3$\times$10$^{-6}$, see $\S$2.3

In addition, three of the nearby bright stars have repeat observations separated by a year 
or more. HR4540 was observed in 2004, 2005 and 2006.
HR5854 was observed in 2004 and 2005 and HR4534 was observed in 2005 and 2006.
HR4540 shows no variation between the three epochs within 1$\sigma$ uncertainties
of $\approx$2.2$\times$10$^{-6}$. HR5854 shows differences between 2004 and 2005 of
5.2$\pm2.0$$\times$10$^{-6}$ in Q/I and 5.1$\pm2.2$$\times$10$^{-6}$ in U/I.
HR4534 shows differences of between 2005 and 2006 of 3.2$\pm1.7$$\times$10$^{-6}$ in Q/I
and -8.1$\pm1.5$$\times$10$^{-6}$ in U/I. Most of these apparent differences have $<$3$\sigma$
significance so we cannot say with certainty whether real variations in the polarisation
have been detected with such a small sample of measurements. We note that the apparently
significant change in the U/I measurement for HR4534 would appear smaller and have a larger
uncertainty if we had not excluded two outlying readings from the average 2006 measurement given
in Table 2.

It is therefore unclear whether the polarisation of nearby stars is stable enough on timescales 
of years to define very low polarisation standards for general use by the community.
If this were possible it would allow high precision polarimeters to be used at equatorially 
mounted telescopes (for which the TP cannot be separated from the polarisation of the light 
source) and allow them to be used at telescopes with alt-azimuth mounts without the need for 
such lengthy calibration observations.

The apparent stability of the polarisation measurements within each observing run suggests that 
neither normal stars nor the interstellar medium in the local ionised bubble produce variable
polarisation on timescales of a few days and that at most only modest changes occur on timescales of 
years. The typical relative stellar motions of 10 kms$^{-1}$ in the solar neighbourhood cause the 
dust column sampled by the line of sight through the interstellar medium to change on a timescale 
of a few days for nearby stars. Assuming that the observed polarisations are due to dichroic
extinction by interstellar dust grains aligned with the local magnetic field, we can 
therefore conclude the dust density, degree of aligment and magnetic field direction is 
typically uniform on spatial scales up to 6$\times$10$^{9}$m (corresponding to the movement
of the dust column in a week) and that large changes do not occur on scales up to $\sim$1~AU
(corresponding to the motion in a year).

\subsection{New detectors}

{\sc Planetpol} employs Avalanche Photodiode Detectors (APDs) to detect the incident radiation.  
Subsequent to the observations reported in this paper, it was found (during brief observations
with the same telescope in November 2006) that the S2383 APDs used for these observations and in 
Hough et al.(2006) (detector size 1mm, gain 100) can be usefully replaced by S2384 APDs (detector 
size 3mm, gain 50) for stars brighter than I $\sim$ 8.5 mag., using a 4-m class telescope with 
no filter. Although the intrinsic noise of the S2384 APD is $\sim$ 7.5 times that of the S2383, 
the excess noise produced by the avalanche process, which is
poportional to the photon noise, is somewhat lower. The overall signal to noise ratio, 
as measured by the internal errors on individual measurements, is improved by ~40\% for bright 
stars. This is equivalent to a factor of two in observing speed. This reduces the 
photon noise multiplicative factor of $\sim$2.5 reported by Hough et al.(2006) to 
$\sim $1.8. The two detector assemblies can easily be interchanged as the Fabry lenses, 
different for the two APDs, are an integral part of these assemblies.

\section{Conclusions}

The strong upper limit for 55 Cnc e, $A_G$$<$0.13$(R/1.2 R_{Jup})^{-2}$$p_m^{-1}$, is hard to explain in 
terms of a low albedo. At the time the observations were planned there were no available 
calculations or data concerning the likely size of very strongly irradiated Neptune mass planets. 
Subsequent calculations by Baraffe et al.(2006) suggest that
while a wide range of sizes (0.4~$R_{Jup}$ to $\sim 1~R_{Jup}$ are possible, it is likely
that hot Neptune planets will usually have radii $\le 0.6 R_{Jup}$. The larger radii can occur
for hydrogen and helium dominated Neptune mass planets but these may evaporate on a relatively 
short timescale. Unfortunately the physics of the evaporation process is still not clearly
established. The
sole known transiting hot Neptune, GJ436b, has a radius between 0.38 and 0.44~R$_{Jup}$ 
(Gillon et al.2007a; 2007b; Deming et al.2007; Bean et al.2008; Torres et al.2007), 
though this may not be a useful comparison planet owing to the much weaker irradiation from its
M2V star. Our non-detection of 55 Cne e is most easily explained by a relatively small planet.
The data allow us to rule out a Jupiter sized planet with even a low geometric albedo and 
also rule out a 0.8$~R_{Jup}$ planet with a moderate albedo. However, we caution that it
is possible for planets to have very low geometric albedos. For example the radiative transfer
could be dominated by very small dust particles with a very small single scattering albedo. Also,
if there is no high altitude dust layer, strong atomic and molecular absorption features 
such as the KI resonant line at 770 nm could reduce the geometric albedo in the {\sc Planetpol}
passband. Furthermore, Hood et al.(2008) have shown that 3D structure in planetary atmospheres
can reduces the geometric albedo by a large fraction, relative to homogeneous atmospheres such as those
modelled in this work. Hence a planetary radius as high as $R$=1.2~$R_{Jup}$ cannot be entirely ruled out.

We conclude from the upper limit for $\tau$ Boo b that this is a fairly dark planet in our 590-920 nm
passband. This extends the similar limits that were found by Charbonneau et al.(1999) and Leigh et 
al.(2003) at shorter wavelengths by searching for Doppler shifted absorption lines in the stellar spectrum. 
A non-detection at an upper limit of $A_G$$<$0.37$(R/1.2 R_{Jup})^{-2}$$p_m^{-1}$ is not 
particularly surprising in view of the possible absorption mechanisms described above.
The most plausible example model of Seager et al.(2000), in which the scattering is dominated by a 
mixture of three types of 0.1~$\mu$m grains, gave $A_G=0.175$. Nonetheless, the data indicate
that $\tau$~Boo b is less reflective than Jupiter and Saturn, which have $A_G$$\approx$0.5
throughout most of the {\sc Planetpol} passband (Karkoschka et al.1994). This is of course
subject to the assumption that the scattering particles have Rayleigh-like polarisation behaviour,
i.e. the scattering is dominated by molecules or by dust particles 
$\la$0.1~$\mu$m in radius.
Both our data and the {\it MOST} data indicate that this planet is not an ideal target for the highest 
precision studies.

Our multiple scattering model atmosphere calculations indicate that the large amplitude periodic
polarisation signal from the HD189733 system that was reported by Berdyugina et al.(2008) cannot
be explained in terms of reflected light from the planet HD189733b. If the observations are confirmed
it would be important to consider the possible contribution of star spots to the polarisation of the system,
given that HD189733 is an active star with much larger photometric variations than $\tau$ Boo or
55 Cnc, e.g. Winn et al.(2007).

The measurements for nearby stars with low polarisation are stable on timescales of a week.
It is not yet clear whether they are stable at the 10$^{-6}$ level on timescales of a year.
Finally, we note that these results do not represent the full potential of the instrument.
The new detectors offer a 40\% improvement in sensitivity, so observations of stable
bright hot Jupiter systems such as $\upsilon$~And and 51 Peg would offer a good chance of 
a successful detection. The new class of very hot Jupiters recently detected by transit
surveys around fainter stars may also offer good prospects for observation. Transiting planets, 
which have $i$$\approx$90$^{\circ}$, display the full range of phase angles (0-180$^{\circ}$) during the
course of an orbit, thereby offering the maximum possible amount of information. For the purpose of simple
detection, there is less need to sample the full phase curve since the most likely timing of maximum 
polarisation is predictable if we know that $i$$\approx$90$^{\circ}$ and assume that the scattering
phase function is not highly aniostropic. However, the expected amplitude of the periodic polarisation 
changes in Stokes Q/I and U/I is slightly smaller at $i$=90$^{\circ}$ than for smaller inclinations, and we 
caution that  planets in such small orbits might induce variable polarisation on their central star due to 
star spots.

\section{Acknowledgments}

We wish to thank the staff of the William Herschel Telescope and the members of the 
PPARC grant panel who supported this project. {\sc Planetpol} was partially funded
by a 100k grant from PPARC, the predecessor to the UK Science and Technology Facilties
Council. We also thank Matt Giguere, Debra Fischer and Geoff Marcy for providing orbital phase 
information for 55 Cnc e and $\tau$ Boo b.


\end{document}